\documentclass[twocolumn,tighten,times]{aastex63}

\usepackage{xspace}
\usepackage{float}
\usepackage[normalem]{ulem}

\newcommand{\Ss}{{\tt S06}\xspace}
\newcommand{\Ns}{{\tt S0}\xspace}

\def\lesssim{\mathrel{\hbox{\rlap{\hbox{\lower4pt\hbox{$\sim$}}}\hbox{$<$}}}}
\def\gtrsim{\mathrel{\hbox{\rlap{\hbox{\lower4pt\hbox{$\sim$}}}\hbox{$>$}}}}
\newcommand{\bea}{\begin{eqnarray}}
\newcommand{\eea}{\end{eqnarray}}

\newcommand{\harm}{{\tt Harm3d}\xspace}
\newcommand{\bothros}{{\tt Bothros}\xspace}

\newcommand{\msun}{M_\odot}
\newcommand{\mdotedd}{\dot{M}_{\rm Edd}}

\newcommand{\hz}{{\rm Hz}}

\newcommand{\xx    }[1]{x^{\left(#1\right)}}

\newcommand{\xone  }{\xx{1}}
\newcommand{\xtwo  }{\xx{2}}
\newcommand{\xthree}{\xx{3}}

\def\lambdabar{
\relax
\bgroup
\def\@tempa{\hbox{\raise.73\ht0
\hbox to0pt{\kern.25\wd0\vrule width.5\wd0
height.1pt depth.1pt\hss}\box0}}%
\mathchoice{\setbox0\hbox{$\displaystyle\lambda$}\@tempa}%
{\setbox0\hbox{$\textstyle\lambda$}\@tempa}%
{\setbox0\hbox{$\scriptstyle\lambda$}\@tempa}%
{\setbox0\hbox{$\scriptscriptstyle\lambda$}\@tempa}%
\egroup
}

\usepackage{soul}

\received{2021 December 17}
\revised{2022 February 16}
\accepted{2022 February 17}
\published{2022 April 4}
\submitjournal{ApJ}

\shorttitle{Electromagnetic Emission from SMBBHs}
\shortauthors{Guti\'errez et al.}
\graphicspath{{./}{figures/}}

\begin{document}

\title{Electromagnetic Signatures from Supermassive Binary Black Holes Approaching Merger}

\correspondingauthor{Eduardo M. Guti\'errez}
\email{emgutierrez@iar.unlp.edu.ar}

\author[0000-0001-7941-801X]{Eduardo M. Guti\'errez}
\affiliation{Instituto Argentino de Radioastronom\'ia (IAR, CCT La Plata, CONICET/CIC), C.C.5, (1984) Villa Elisa, Buenos Aires, Argentina}
\affiliation{Center for Computational Relativity and Gravitation, Rochester Institute of Technology, Rochester, NY 14623, USA}

\author[0000-0002-5427-1207]{Luciano Combi}
\affiliation{Instituto Argentino de Radioastronom\'ia (IAR, CCT La Plata, CONICET/CIC), C.C.5, (1984) Villa Elisa, Buenos Aires, Argentina}
\affiliation{Center for Computational Relativity and Gravitation, Rochester Institute of Technology, Rochester, NY 14623, USA}

\author[0000-0003-3547-8306]{Scott C. Noble}
\affiliation{Gravitational Astrophysics Laboratory, NASA Goddard Space Flight Center, Greenbelt, MD 20771, USA}

\author[0000-0002-8659-6591]{Manuela Campanelli}
\affiliation{Center for Computational Relativity and Gravitation, Rochester Institute of Technology, Rochester, NY 14623, USA}

\author[0000-0002-2995-7717]{Julian H. Krolik}
\affiliation{Department of Physics and Astronomy, Johns Hopkins University, Baltimore, MD 21218, USA}

\author[0000-0002-4882-5672]{Federico L\'opez Armengol}
\affiliation{Center for Computational Relativity and Gravitation, Rochester Institute of Technology, Rochester, NY 14623, USA}

\author[0000-0001-9072-4069]{Federico Garc\'ia}
\affiliation{Instituto Argentino de Radioastronom\'ia (IAR, CCT La Plata, CONICET/CIC), C.C.5, (1984) Villa Elisa, Buenos Aires, Argentina}

\begin{abstract}
We present fully relativistic predictions for the electromagnetic emission produced by accretion disks surrounding spinning and nonspinning supermassive binary black holes on the verge of merging. We use the code \bothros to post-process data from 3D General Relativistic Magnetohydrodynamic (GRMHD) simulations via ray-tracing calculations. These simulations model the dynamics of a circumbinary disk and the mini-disks that form around two equal-mass black holes orbiting each other at an initial separation of $20$ gravitational radii, and evolve the system for more than $10$ orbits in the inspiral regime.  We model the emission as the sum of thermal blackbody radiation emitted by an optically thick accretion disk and a power-law spectrum extending to hard X-rays emitted by a hot optically thin corona. We generate time-dependent spectra, images, and light curves at various frequencies to investigate intrinsic periodic signals in the emission, as well as the effects of the black hole spin. We find that prograde black hole spin makes mini-disks brighter since the smaller innermost stable circular orbit angular momentum demands more dissipation before matter plunges to the horizon. However, compared to mini-disks in larger separation binaries with spinning black holes, our mini-disks are less luminous: unlike those systems, their mass accretion rate is lower than in the circumbinary disk, and they radiate with lower efficiency because their inflow times are shorter. Compared to a single black hole system matched in mass and accretion rate, these binaries have spectra noticeably weaker and softer in the UV. Finally, we discuss the implications of our findings for the potential observability of these systems.

\end{abstract}


\section{Introduction} \label{sec:intro}

In the past few years, the LIGO-Virgo Collaboration has detected $\sim 90$ gravitational wave (GW) signals produced by the merger of stellar-mass compact objects \citep{GWdetect, abbott2021gwtc}. Among these, a collision of two neutron stars was accompanied by the detection of electromagnetic (EM) radiation across the entire EM spectrum, marking the most significant multimessenger astrophysical event to date \citep{abbott2017gw170817}. Although the majority of GW detections have been associated with binary black hole (BH) mergers, no reliable EM signal has been detected from these events so far. This is unsurprising for stellar-mass binary BHs, as most of them would merge in a low-density environment \citep{perna2018binary}. On the other hand, and according to our current understanding of galaxy evolution, supermassive binary black holes (SMBBHs) would form and evolve in gas-rich environments, specifically at the centers of merged galaxies \citep{Begelman1980,Escala2004, Escala2005, Merrit2004, Merritt2006,  Springel2005, Dotti2007, Mayer2007, volonteri2010formation, Dotti2009b, Shi2012,SesanaKhan2015,Mirza2017,Khan2019, Tiede+2020, Chapon2013}. Additionally, since accretion at the Bondi radius scales as $M^2$ for fixed conditions in the neighboring interstellar medium, SMBBH systems may accrete copious amounts of gas, forming a disk if the matter has enough angular momentum. SMBBHs may thus emit EM waves \textit{and} GWs, as they inspiral and finally coalesce \citep{bogdanovic2021electromagnetic}.

GWs from SMBBHs have frequencies ranging from nanohertz to millihertz frequencies. At the lower end of this frequency range, they are targets of current Pulsar Timing Array experiments \citep{Reardon2015x,EPTA2016,Nanograv2020}; at the higher-end, they are targets for future space-based observatories such as the Laser Interferometer Space Antenna \citep[LISA,][]{LISA2017}\footnote{\url{https://www.elisascience.org/}}. Although direct detection of GWs from SMBBHs is still a decade away, EM emission may aid in the discovery of SMBBHs using current large-scale surveys such as SDSS-V, eROSITA, and upcoming facilities like the Vera Rubin Observatory. Identifying these systems via their EM emission will also assist GW missions in constraining population estimates to find real-time counterparts. Since SMBBHs are expected to have a sufficient amount of available gas for accretion, their luminosity should be, a priori, comparable to that of normal AGNs. The major difficulty is differentiating SMBBHs from conventional AGNs powered by single BHs. A small portion of the total AGN population should correspond to SMBBHs \citep{kelley2019massive, Krolik2019}, but at present, although there are several candidates \citep{valtonen2008massive, dorazio2015nature,graham2015systematic,hu2020spikey,ONeill2022}, there are no confirmed detections.

To accurately predict distinct EM signatures from SMBBHs, we must first understand how they behave. This is a difficult problem that requires solving the nonlinear dynamics of the plasma, magnetic fields, and radiation, coupled with a dynamical spacetime. To accomplish this, we must rely on numerical simulations that can handle all of these physical ingredients.

Nonetheless, some results are well established. For example, according to 2D-viscous simulations \citep{MM08,DOrazio13,Farris2014a,Farris2014b,DOrazio2016,Munoz2016,Miranda2017,
Derdzinski2019,Munoz2019,Moody2019,Mosta2019,Duffell2020,Zrake2021,Munoz2020a,Munoz2020b,
Tiede2020,Derdzinski2021}, Newtonian 3D-MHD simulations \citep{Shi2012}, and 3D-GRMHD simulations \citep{Noble12, Zilhao2015, lopezarmengol2021, noble2021, Bowen2018, Bowen2019, Gold14, Gold2014b, Farris11, paschalidis2021minidisk,cattorini2021,giacomazzo2012}, binaries with a mass ratio $q:=m_1/m_2 > 0.04$ carve an eccentric cavity around their center of mass, whose mean radius is roughly two times the binary separation. More importantly, simulations of this sort revealed that contrary to previous expectations \citep{Pringle1991x, Milosavljevic2005, Kocsis:2011dr}, binary torques do not completely halt mass accretion \citep{MM08,Shi2012,Farris14,Shi2015}.
For mass ratios of $q>0.1$, the circumbinary disk (CBD) develops an $m=1$ density mode on its inner edge, known as the lump, that modulates the mass accretion onto the binary \citep{Noble12, Shi2012, Farris14, noble2021}.

The gas enters the cavity as a thin ballistic stream \citep{Shi2015} and forms mini-disks around the black holes. The mini-disks' properties depend on the ratio $r_{\rm ISCO}/r_{\rm trunc}$, where $r_{\rm ISCO}$ is the radius of the innermost stable circular orbit (ISCO) and $R_{\rm trunc}$ is the mini-disk's tidal truncation radius, which is roughly $\sim 0.35-0.4 r_{12}$ \citep{Bowen17}; here, $r_{12}$ is the binary separation. At short separations, the inflow time in the mini-disks can be shorter than the beat frequency between the orbital frequencies of the lump and the binary, and thus their mass and accretion rate go through a filling-depletion cycle \citep{Bowen2018, Bowen2019}. At separations large enough for the mini-disk's inflow time to be longer than the beat frequency, however, any modulation of thermal disk radiation due to supply rate modulation is strongly suppressed.  In this regime, the lump-driven accretion rate periodicity can be seen only when the associated radiation is created when the stream strikes a mini-disk \citep{Sesana2012,Roedig2014}.

If the BHs have prograde spin, the smaller angular momentum of an ISCO means that accreting matter must lose more angular momentum. The greater time needed to do so increases the mass resident in the mini-disks for a fixed accretion rate \citep{paschalidis2021minidisk, Combi2021b}. Spin can also affect the streams from the CBD via frame-dragging \citep{lopezarmengol2021}. It can also power jets through the Blandford--Znajek mechanism as seen in both force-free \citep{palenzuela2010dual, moesta2012detectability} and ideal GRMHD \citep{Kelly2017, cattorini2021, Combi2021b, paschalidis2021minidisk} mini-disk simulations.

Finally, when the SMBBH coalesces due to GW emission \citep{Campanelli:2006uy}, other EM signatures might appear, although these predictions are much less robust than those listed above.  Possibilities include: jet emission interruption \citep{Shoenmakers2000, Liu2003}, prompt Eddington-limited thermal radiation \citep{Krolik10}, and a variety of imprints due to black hole recoil \citep{Campanelli:2007apj,SK08,volonteri2008off,ONeill09,Rossi09, blecha2016recoiling}.

Although much work remains to be done to fully understand key aspects of the system, these findings clearly show that (a) SMBBHs can accrete significantly, and (b) at close separation, the accretion process has several quasi-periodicities associated with the binary motion. This variability will then be reflected in the EM emission, which may uniquely indicate the presence of a binary system \citep{dAscoli2018}.
Aside from the accretion flow's intrinsic variability, relativistic effects such as Doppler shifting \citep{dorazio2015nature,charisi2018testing} and self-lensing \citep{Dorazio2018, kelly2021electromagnetic, ingram2021self, davelaar2021a, davelaar2021b} may provide additional signatures for detecting SMBBHs when the binary separation is small.

The EM emission from SMBBHs depends on how photons move and interact with the plasma. This is difficult to model self-consistently and has thus far been addressed only for single BHs in specific regimes \citep{SDavis2005,ZhuDavis2012,Kinch2021}. An intermediate step consists of implementing some prescription for how the gas radiates, such as a cooling function \citep{Noble09}, and then tracking the photons as they travel through the highly dynamical spacetime. \citet{dAscoli2018} used GRMHD simulations from \citet{Bowen2019}, that simulate mini-disk accretion onto nonspinning SMBBHs, using the metric approach in \citet{Mundim2014}, and implemented a fully relativistic ray-tracing calculation using the code \bothros \citep{Noble07}.

In this paper, we extend the work done by \citet{dAscoli2018} to a much longer simulation and include a new simulation with spinning black holes performed by \citet{Combi2021b}, which uses a new metric approach introduced in \citet{Combi2021a}. The longer duration of the runs allows the system to reach a relaxed state, which enables us to calculate the first detailed fully relativistic light curves of an SMBBH system approaching merger. We seek to answer questions such as: what type of spectra do SMBBHs produce and how do they differ from those of single AGNs? How do mini-disks radiate and how closely do they resemble standard disks around single BHs? Are there unique features in the time variability of the emission? What are the prospects for observing these systems?

The remainder of the article is organized as follows: in Section \ref{sec:methodology}, we describe the GRMHD simulations we used and the methodology we adopted to obtain the scientific products of the work, which we present in Section \ref{sec:results}. These consist of images, spectra, and light curves. In Section \ref{sec:discussion}, we discuss the main implications of our findings and address the above-mentioned questions. Finally, we present our conclusions in Section \ref{sec:conclusions}.

\section{Methodology} \label{sec:methodology}

We compute images, spectra, and light curves by performing ray-tracing calculations with data from GRMHD simulations. Throughout the work, we use geometrized units, $G=c=1$, to describe simulation data and CGS units to present EM observables.

\subsection{GRMHD Simulations} \label{subsec:GRMHD}
 
We use data from two GRMHD simulations: one with aligned spins, hereafter \Ss, performed in \citet{Combi2021b}, and the other with zero spins, hereafter \Ns, performed in \citet{Bowen2018, Bowen2019}. In both simulations, the GRMHD equations for the plasma are evolved using the finite-volume code \harm with the same initial data and numerical setup, with the only difference being the metric spacetime. 

In \Ns, the plasma evolves on top of a binary black hole spacetime represented by a semi-analytical approximate metric constructed using a matching technique \citep{Mundim2014}, and in which the BHs have zero spins. In \Ss, the spacetime is represented by a different approximate metric, built through a superposition, in which BHs have aligned spins of $\chi_i \equiv J_i/m_i^2 = 0.6$, where $J_i$ and $m_i$ are the angular momentum and mass of the $i$th black hole. \citet{Combi2021a} demonstrated that these two spacetime metrics---in both of which the binary separation is $a=20M$ in the initial state---are physically equivalent, and the only meaningful difference in the simulations is the spin of the black holes.

The initial data for the CBD are taken from a snapshot in \citet{Noble12} in which the system evolved for $50,000M$, reaching a relaxed state in which an $m=1$ overdensity, known as the \textit{lump}, orbits around the inner edge of the circumbinary cavity. Computing the time-average of the accretion rate, we found that the CBD is in excellent inflow equilibrium up to a distance $r\sim 50 M$, a little bit outside its inner edge.  From there to $\simeq 100M = 5a$, the mean accretion rate increases slowly, rising by $\simeq 50\%$.
For this reason, the lowest-frequency portions of the CBD disk spectrum contribution are somewhat stronger than they would be if we had a genuine inflow equilibrium.  However, because these low frequencies are not in a part of the spectrum we use (see Section 3), this inconsistency does not affect any of our conclusions.

The black holes are separated by a distance of $r_{12}=20M$, where $M=m_1+m_2$ is the total mass of the system. This data is interpolated into a new grid that includes the BHs and two mini-disks on quasi-equilibrium and then cleaned of magnetic divergences; see \citet{Bowen2018} for more details. The simulations use outflow boundary conditions on the radial boundaries, reflective, axisymmetric boundary conditions at the polar axis cutout, and periodic boundary conditions on the azimuthal coordinates, and follows the black hole inspiral using post-Newtonian trajectories. 

\begin{figure}
    \centering
    \includegraphics[width=1.\linewidth]{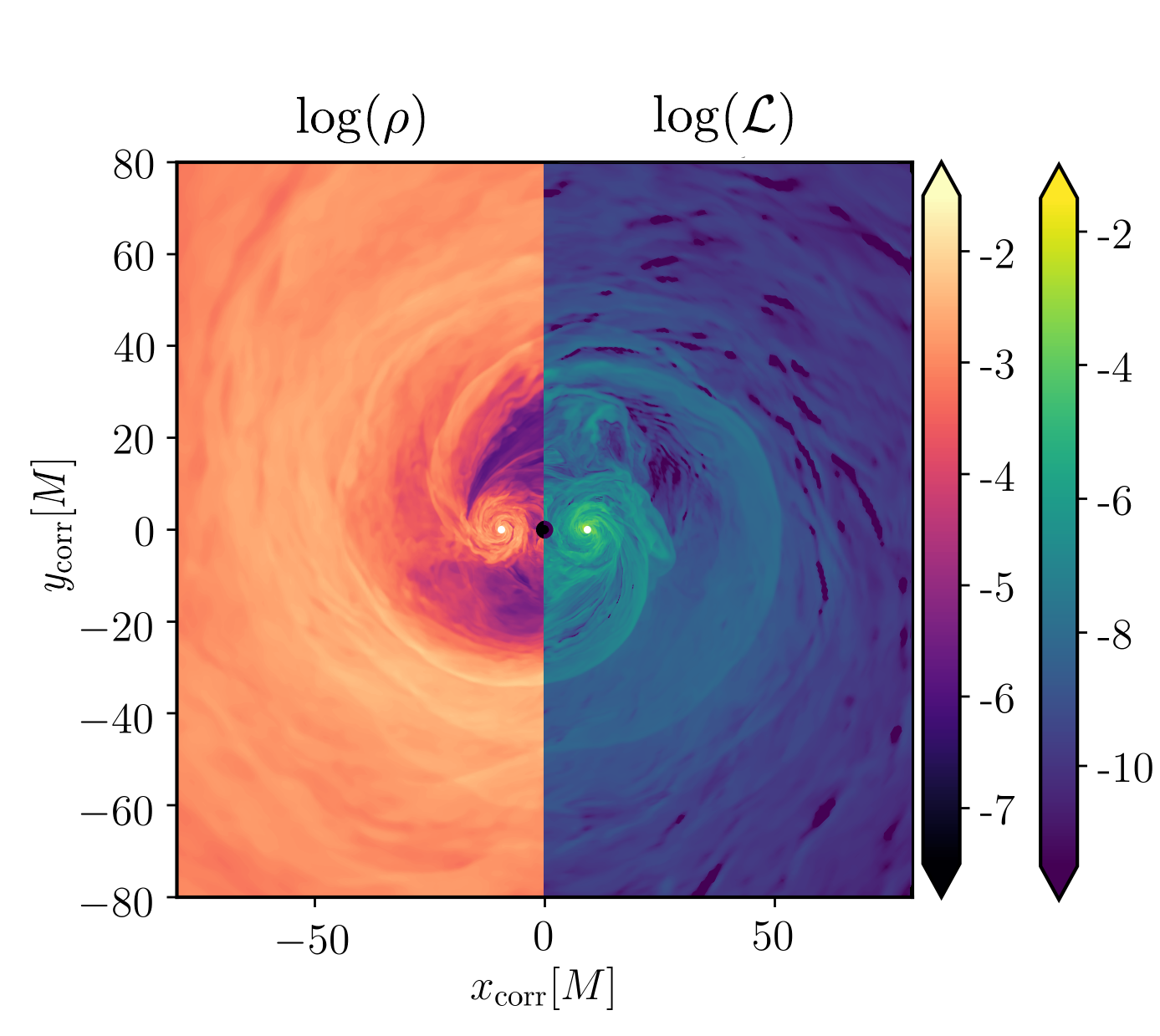}
    \caption{Equatorial slice of the rest-mass density $\rho$ (left panel) and the cooling function $\mathcal{L}_{\rm c}$ (right panel) for \Ss at $t= 3290 M$}
    \label{fig:eqrhocool}
\end{figure}

After a transient of  $\sim 3$ orbits, the mini-disks settle into a slowly-evolving limit cycle: they fill and drain, but with slowly-changing maximum and minimum masses. This cycle is quasi-steady after the transient, as can be seen in Figure~3 of \cite{Bowen2019} and Figure~4 of \cite{Combi2021b}. In Figure~\ref{fig:eqrhocool} we show an equatorial view of the density and cooling function of the GRMHD simulation. Simulation \Ss (\Ns) lasts for a total of $15$ ($12$) orbits, ending with the black holes at a separation of $\sim 16.7M$ ($17.3M$).

In \harm, the energy-momentum conservation equations have a loss-term that radiates away orbital energy dissipated into heat:
\begin{equation}
\nabla_{\mu} T^{\mu \nu} = \mathcal{L}_c u^\nu,
\end{equation}
The cooling function is
\begin{equation}
    \mathcal{L}_{\rm c}= \frac{\rho \epsilon}{t_{\rm cool}} \left ( \frac{\Delta S}{S_0} + \left | \frac{\Delta S}{S_0} \right | \right),
\end{equation}
where $\rho$ is the rest-mass density, $\epsilon$ is the specific internal energy, $\Delta S \equiv S - S_0$, $S$ is the specific entropy, $S_0=0.01$ is the initial specific entropy, and the cooling timescale $t_{\rm cool}$ is defined as
\begin{equation}
    t_{\rm cool}(r) = \left\{ 
    \begin{array}{lll}
    2\pi \left ( r + M \right )^{3/2} / \sqrt{M}, & {\rm if\ } r \ge 1.5 a, \\
    2\pi r_{{\rm BL,}i}^{3/2}/\sqrt{m_i}, & {\rm if\ } r_i \le 0.45 a, \\
    2\pi \left( 1.5a + M\right)^{3/2}/\sqrt{M}, & {\rm otherwise}.
    \end{array}
    \right.
    \label{eq:t_cool}
\end{equation}
In the equation above, $r_{{\rm BL,}i}$ is the Boyer--Lindquist radial coordinate in the co-moving frame of the $i$th BH, and $m_i$ is the BH's mass. Fixing the target entropy $S_0$, we control the temperature of the fluid and thus its height-radius aspect to $h/r \approx 0.1$. 

The cooling term assures that all of the energy dissipated is radiated, and the distribution in space and time of this radiation is very close to the dissipation distribution.  In codes modeling ideal MHD, dissipation is a numerical grid-scale effect, but because physical dissipation tends to be enhanced by sharp gradients, whether of velocity or magnetic field, grid-scale dissipation mimics physical dissipation.

\subsection{Radiative Transfer} \label{subsec:radiative_transfer}

We use the code \bothros (\citealt{Noble07, dAscoli2018}) to calculate the EM emission from our GRMHD simulations. Our methodology follows the one used in \citet{dAscoli2018}. We use a camera-to-source approach, in which photons are launched from the camera going backward in time toward the source. We also perform the integration using the fast-light approximation, which assumes that photons move much faster than the plasma and the spacetime itself\footnote{The validity of this approximation for SMBBHs at the separations involved in this work ($\sim 20~M$) was discussed in detail in \citet[][see Section 4.4]{dAscoli2018}.}.

\begin{figure}
    \centering
    \includegraphics[width=1.\linewidth]{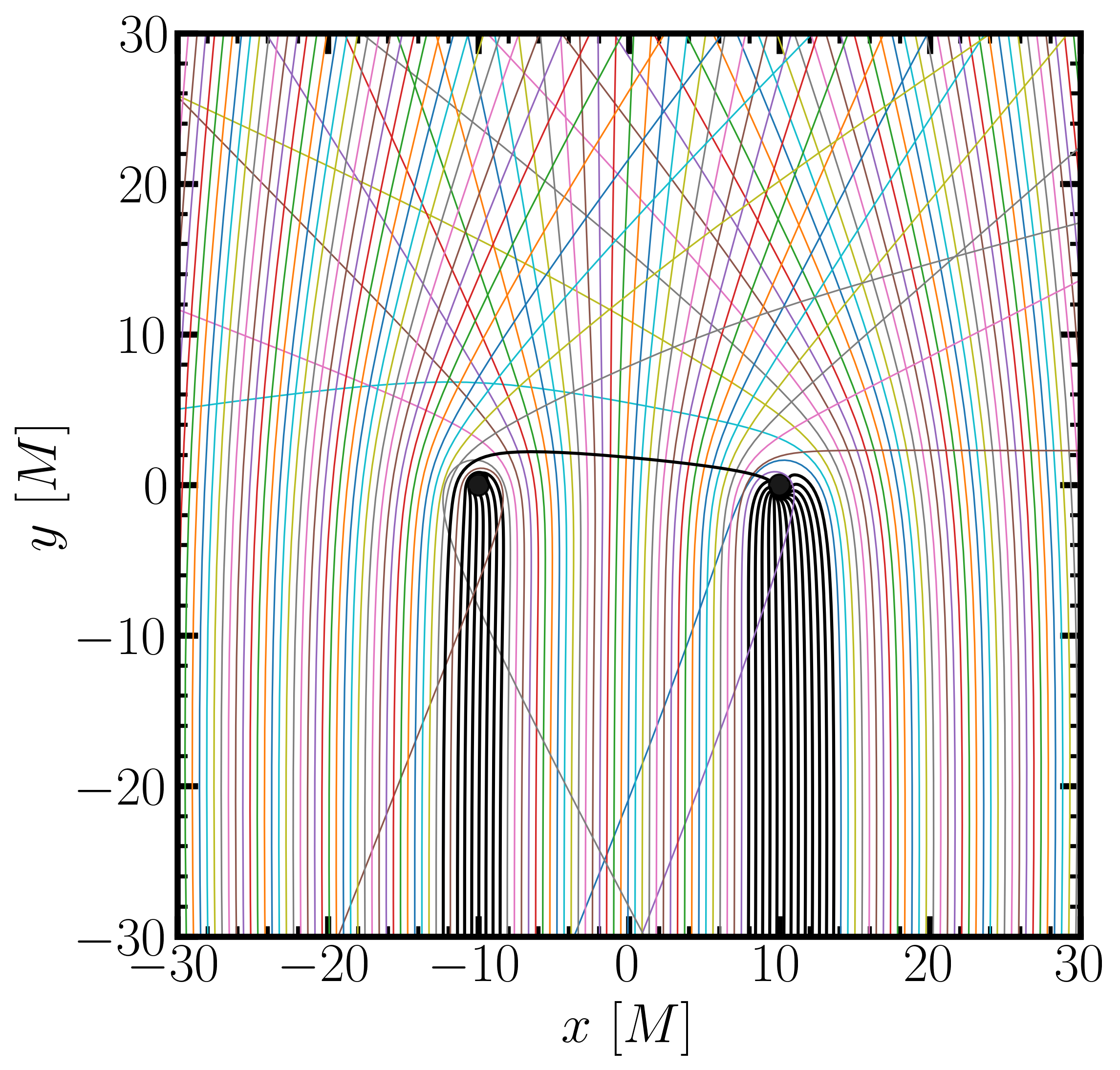}
    \caption{Geodesic paths for a fixed time. The black lines represent those geodesics arising from the event horizon.}
    \label{fig:geodesics}
\end{figure}

In Figure~\ref{fig:geodesics}, we show the paths of a sample of geodesics launched from a camera located on the equatorial plane of the binary system at ($x=0$, $y=-1000M$). We distinguish the geodesics that arise from the event horizon of the holes with dark solid lines; at infinity, these define the black hole shadows\footnote{In emission models of single BHs with very low accretion rate, there is typically a flux depression present in the apparent image. The size of this ``hole'' is model-dependent, but, in general, it coincides with the apparent position of the event horizon. (see, e.g., \citealt{bronzwaer2021, Gralla2021}).}.

After calculating the full sample of geodesic paths for a given time (once per pixel), we integrate the radiative transfer equation along the geodesics to produce maps of specific intensity $I_{\nu_\infty}$ at the camera for a given observed frequency $\nu_\infty$. From these intensity maps, we then calculate flux spectra and light curves. To faithfully recover the flux spectrum for each snapshot, we use a resolution of $800 \times 800$ pixels ($\sim 7~{\rm pixels}/M$) for the region $r<60M$ and $500 \times 500$ pixels ($\sim 2~{\rm pixels}/M$) for the region $60M < r < 150M$. This resolution is sufficient to compute the flux to a precision of $\sim 1 \%$.

To obtain $I_{\nu_\infty}$ at each pixel, we first calculate the optical depth along the geodesic and determine whether the disk's photosphere, defined as the loci of points where $\tau=1$, is reached or not. If $\tau<1$ along the geodesic, then we start the radiative transfer integration from the last point in the geodesic (either at the end of the simulation domain or at the event horizon of one of the black holes) with the initial condition $I_0=0$. On the other hand, if the photosphere is reached, we start the integration from the photosphere with the initial condition $I_0=I_{\rm photosphere}=B_\nu(T_{\rm eff})$. Here, $B_\nu(T)$ is the Planck function, and $T_{\rm eff}$ is the photosphere's effective temperature at the point where the given geodesic intersected it. Placing this initial condition is equivalent to assuming that the photosphere radiates a blackbody spectrum at the local value of the effective temperature.

The effective temperature is calculated using the Stefan--Boltzmann law from the radiative cooling flux ${\cal F}$:
\begin{equation}
    T_{\rm eff}= \left( {\cal F} / \sigma \right)^{1/4},
\end{equation}
where $\sigma$ is the Stefan--Boltzmann constant. In turn, the radiative flux is calculated by integrating vertically the cooling function inside the photosphere:
\begin{equation}
    {\cal F} = \frac{1}{2} \int_{\tau > 1} \mathcal{L}_{\rm c}(z) dz,
\end{equation}
where the factor $1/2$ comes from the fact that the disk has two surfaces through which to cool.

Regardless of the initial point, the radiative transfer equation is solved only in the flow's optically thin region, referred to as the corona.
To perform the integration, we must assume an emission and absorption model. Electrons in AGN coronae reach relativistic temperatures and cool primarily through inverse Compton scattering of low-energy photons from the optically thick disk.
This is a nonlocal process in which a photon may suffer many scatterings before escaping the system. It produces a nonthermal spectrum (power law) with a high energy cutoff, regardless of whether particles follow a thermal energy distribution, which is usually the case. A detailed treatment of this process is very complex and computationally costly (see, however, \citealt{Kinch2020, Kinch2021}). For simplicity, we follow \citet{dAscoli2018} and model the multiple nonlocal Compton scatterings using a local effective emissivity coefficient of the form
\begin{equation}
    j_\nu = \frac{{\cal L}_{\rm c}}{4\pi^{3/2}}\frac{h}{k_{\rm B} T}\left( \frac{h\nu}{k_{\rm B}T} \right)^{-1/2} e^{-\frac{h\nu}{k_{\rm B}T}},
    \label{eq:corona_em}
\end{equation}
that mimics the actual output from the Comptonization.
The normalization is such that $4\pi\int j_\nu d\nu = {\cal L}_{\rm c}$, namely that the bolometric emissivity matches the cooling function at every point, and we fix the dimensionless temperature and spectral index of the corona to typical values observed in luminous single AGNs: $\Theta = k_{\rm B}T/m_{\rm e}c^2=0.2$ and $p=0.5$.

At the temperatures and densities of interest in our scenario, the dominant source of opacity is electron scattering. We assume a gray (frequency-independent) Thomson opacity,
\begin{equation}
    \alpha_\nu = {\rm const.} = \sigma_{\rm T} n_{\rm e},
\end{equation}
where $n_{\rm e}=\rho / m_H$ is the electron number density, and $\sigma_{\rm T}$ is the Thomson cross section.

Finally, after calculating the specific intensity at every pixel for a given time $t$, we integrate over all pixels to obtain the flux:
\begin{equation}
    F_{\nu_\infty} = \int I_{\nu_\infty} (\xi, \eta) d\xi d\eta,
\end{equation}
where ($\xi$, $\eta$) is the pair of angular coordinates at the camera, and we have assumed that this is located sufficiently far from the system so that the rays arrive approximately parallel to each other\footnote{A distance of $r_{\rm cam}=10^3M$ is enough for this approximation to be reasonable: we have checked that the flux changes less than $1 \%$ when choosing $r_{\rm cam}=10^3M$, $10^4M$, or $10^5M$.}. Then, the spectral luminosity is simply $L_{\nu_\infty} = 4\pi r_{\rm cam}^2 F_{\nu_\infty}$.
\vspace{0.5cm}

\subsection{Units Choice}

Our GRMHD simulations ignore the fluid self-gravity, which is not important in this scenario. Thus, there is scale freedom for both the total mass of the binary system and the physical mass density scale of the gas. The latter can be set indirectly using a more easily interpretable quantity such as the accretion rate. 
This means that for each of the two sets of simulation data (spinning and nonspinning), we can investigate various scenarios in which either the total mass $M$, the accretion rate $\dot{M}$, or both change.

The accretion rate is calculated in the simulation in code units (CU) as
\begin{equation}
    \dot{\cal M}(r,t)=-\int_{{\cal S}^2(r)} \rho u^r \sqrt{-g}d\theta d\phi,
    \label{eq:acc_rate}
\end{equation}
where $\rho$ is the fluid mass density, $u^r$ is the radial component of the fluid 4-velocity, $g$ is the determinant of the metric, and ${\cal S}^2(r)$ is a spherical surface of fixed radius $r$ at time $t$ in the harmonic (center of mass) coordinate system. We average $\dot{\cal M}$ over the range $2a<r<4a$ for the initial snapshot; this calculation gives a code-unit value of $\langle \dot{\cal M}\rangle \approx 0.015$. Comparing this to a given assumed physical accretion rate $\dot{M}$, we obtain the mass density scale\footnote{For more details on the conversion from code units to physical units, see Appendix B in \citet{dAscoli2018}. Note, however, that this paper mistakenly describes the code-unit accretion rate for the \citet{Noble12} CBD data we both use as $\approx 0.03$. This correction does not invalidate their results; rather, they correspond to a physical accretion rate half the value assumed there.}.

In both \Ss and \Ns simulations, the flow cools efficiently enough to maintain an average height-to-radius ratio of $h/r \sim 0.1$. This aspect ratio is what is usually expected in quasars, where typical estimates of the accretion rate are $\sim 10^{-1\pm 1}\times$ the Eddington rate.
For rough consistency, we compute the radiative output assuming $\dot{M}=0.25 \mdotedd$, where $\mdotedd \equiv 1.39\times 10^{18}\left( M/\msun \right)~{\rm g~s}^{-1}$ is the Eddington accretion rate with an assumed efficiency of $10\%$. We also set the total mass of the binary to $M=10^6\msun$ as a fiducial value. In Sec. \ref{sec:discussion}, we explore other values for the black hole mass.

We choose a face-on view ($i=0^\circ$) to investigate general features in the emission of these systems that are independent of inclination\footnote{Our photospheric boundary condition assumes an isotropic intensity, so that the total luminosity is $4\pi\times$ the observed luminosity per solid angle.}. At higher viewing angles, the detected radiation would be strongly modulated by relativistic effects such as gravitational lensing and Doppler boosting. These modulations may be a useful tool for identifying SMBBHs through timing analysis, and they will be fully investigated in future work. Note, however, that the emission in the face-on case is still affected by gravitational redshift and transverse Doppler effects.

\section{Results} \label{sec:results}

The dynamics of the CBD and mini-disk accretion onto the relativistic BBH system were thoroughly analyzed in \citet{Combi2021b} and \citet{Bowen2018, Bowen2019}. The most notable feature of these systems is that the lump modulates both the accretion rate and mass of the mini-disks over time. When one BH passes close to the lump, it pulls some of the lump mass into a thin stream that falls almost ballistically into the cavity. This occurs periodically with the beat frequency, $f_{\rm beat}= f_\mathcal{B} -f_{\rm lump} \approx 0.72 f_\mathcal{B}$, where $f_{\rm lump} \approx 0.28 f_\mathcal{B}$ is the orbital frequency of the lump \citep{Noble12, Shi12, lopezarmengol2021}, and $f_\mathcal{B}$ is the orbital frequency of the binary. Matter can fall onto the binary, rather than being torqued up and thrown back at the disk, if its angular momentum with respect to the binary center of mass is $10\%$--$15\%$ less than the orbital angular momentum at the inner edge of the CBD \citep{Shi2015,Tiede2021}. This happens when the thrown-back matter shocks against the CBD, and some material is deflected strongly enough to lose this much in angular momentum.

The infalling fluid has a distribution of specific angular momentum with respect to the BH. A portion of the fluid has enough angular momentum to orbit the BH, while the rest plunges almost directly into the hole \citep{Combi2021b}.
The critical angular momentum distinguishing these two fates, $l_{\rm crit}$, depends on the location of the ISCO, which is a strong function of spin, but is independent of the binary separation $r_{12}$. The maximum size of the mini-disk, on the other hand, is determined by the binary's truncation radius $r_{\rm trunc} \approx 0.35-0.4 r_{12}(t)$. For our close-separation binary, $l_{\rm crit}$ is always less than the $l$ required for a circular orbit at $r_{\rm trunc}$, but only by a factor $\sim O(1)$.  Consequently, even for matter with $l > l_{\rm crit}$, the inflow time is relatively short because only a small diminution in $l$ brings it below $l_{\rm crit}$, while the inflow time for matter arriving with $l < l_{\rm crit}$ is only slightly larger than the radial freefall time. The principal distinction we find between the spinning and nonspinning cases is that $l_{\rm crit}$ is smaller for (prograde) spinning BHs, so that a larger fraction of the accreted mass must orbit for a while in the mini-disk before plunging.

\subsection{Overview of the Emission} \label{subsec:spectrum}
\begin{figure}
    \centering
    \includegraphics[width=0.98\linewidth]{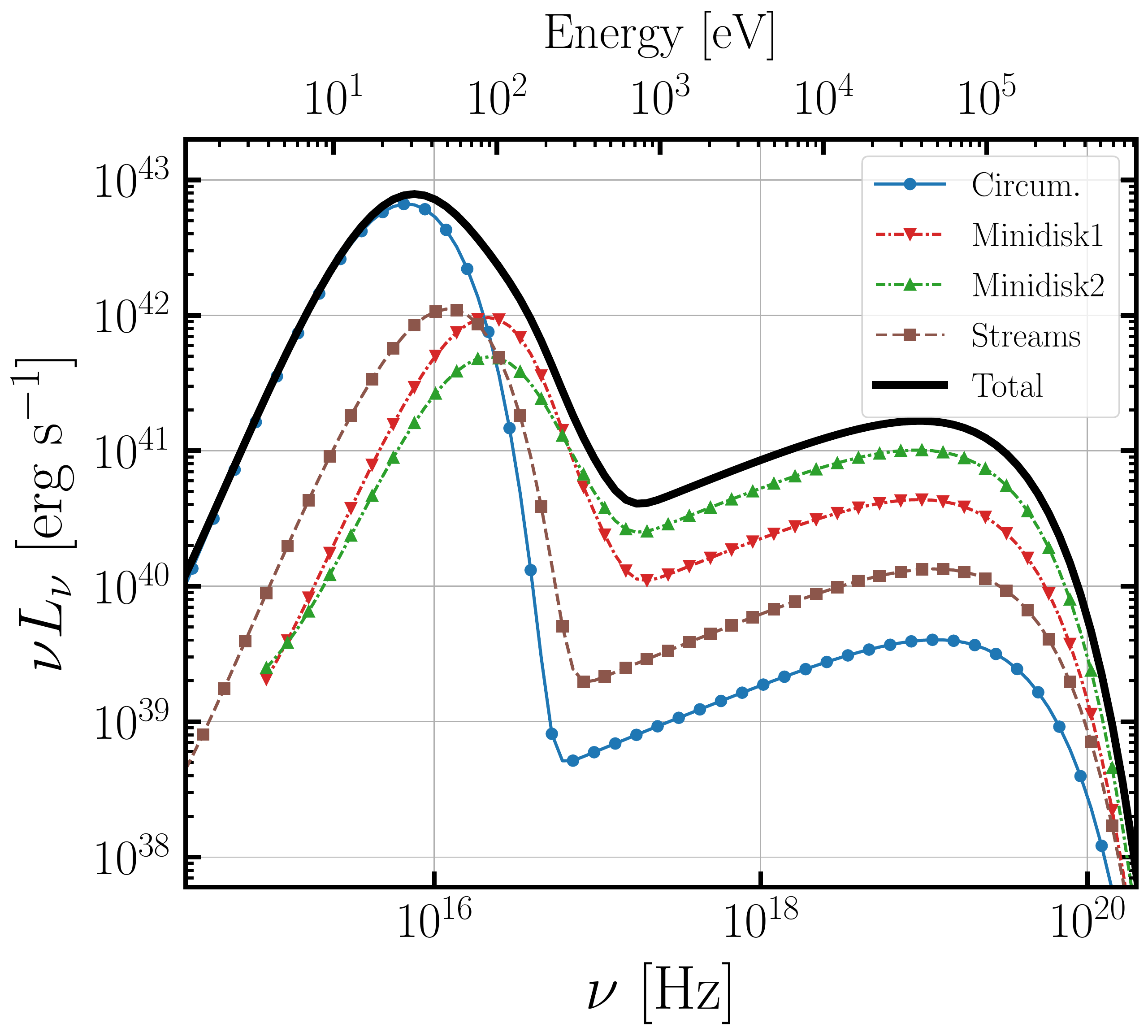}
    \caption{Spectral energy distribution (SED) of the \Ss simulation at $t=3290M$, assuming a total mass of $10^6M_\odot$. The different components are distinguished in the legend. Note that the frequencies of the thermal peaks scale $\propto M_{\rm BH}^{-0.25}$.}
    \label{fig:spectrum}
\end{figure}

The spectrum of the system can be analyzed in terms of its three main contributors: the CBD, the streams, and the mini-disks. In each case, the emission is the sum of two components: a multitemperature blackbody, effectively emitted by the accretion flow's photosphere, and a Comptonized power law extending up to hard X-rays, that reflects the energy dissipation in the corona. We assumed that this emission is produced by the Compton up-scattering of low-energy photons in the optically thin hot corona.
Figure~\ref{fig:spectrum} shows an example of a spectrum for \Ss at a specific time, $t=3290M$. The CBD is defined as the region $r>2r_{12}$, the streams as the region $r_{12}<r<2r_{12}$, and the mini-disks as the regions $r_i<0.45r_{12}$, where $r_i$ denotes the radial coordinate in the reference frame of the $i$th BH. The remaining region is almost devoid of matter and has a negligible contribution; however, it is included in the total emission. The CBD spectrum is depicted by the blue solid line. Its thermal component peaks in the UV band because its inner boundary reaches an effective temperature of $\sim 0.9 \times 10^5$ K; this component dominates the bolometric luminosity of the system. The plasma in the mini-disks reaches higher effective temperatures ($\gtrsim 3.2 \times 10^5~{\rm K}$), and its spectrum peaks in the far-UV. A typical spectrum from \Ns shows the same features described above.

In accretion disks onto single black holes, one would expect the density and temperature to follow a continuous trend down to $\lesssim 1.5 r_{\rm ISCO}$. On the contrary, the region between the CBD and the mini-disks in SMBBHs is usually rather empty and cold. This could result in a notch in the spectrum between the two thermal peaks \citep{Roedig2014}. However, this effect is missing in our spectrum because the mini-disks' luminosity is too low to create the expected rise at short frequencies (as we will show, this condition is specific to very close binaries). An additional minor effect is that the streams' emission lies in the middle of the two thermal peaks, though it always lies below the sum of the CBD's and mini-disk's emission.

In contrast to typical single BH thin disks, the global peak in our spectrum does not correspond to the highest temperatures, reached here in the mini-disks. The mini-disk's low luminosity is caused by a combination of two effects. On one hand, the accretion rate is lower than in the CBD; on the other hand, a portion of the plasma in the mini-disks has low angular momentum; not requiring much angular momentum loss to fall into the black hole, less dissipation takes place within it. Indeed, as shown in \citet{Combi2021b}, while some of the material in the mini-disk can orbit the black hole, another component plunges directly into the hole (see also \citealt{BI2001} for a similar situation potentially occurring in X-ray BH binaries). In Sec. \ref{subsec:eff}, we investigate the relative importance of these two effects in depth.

\subsection{Brightness Maps}

\begin{figure*}
    \centering
    \includegraphics[width=0.99\linewidth]{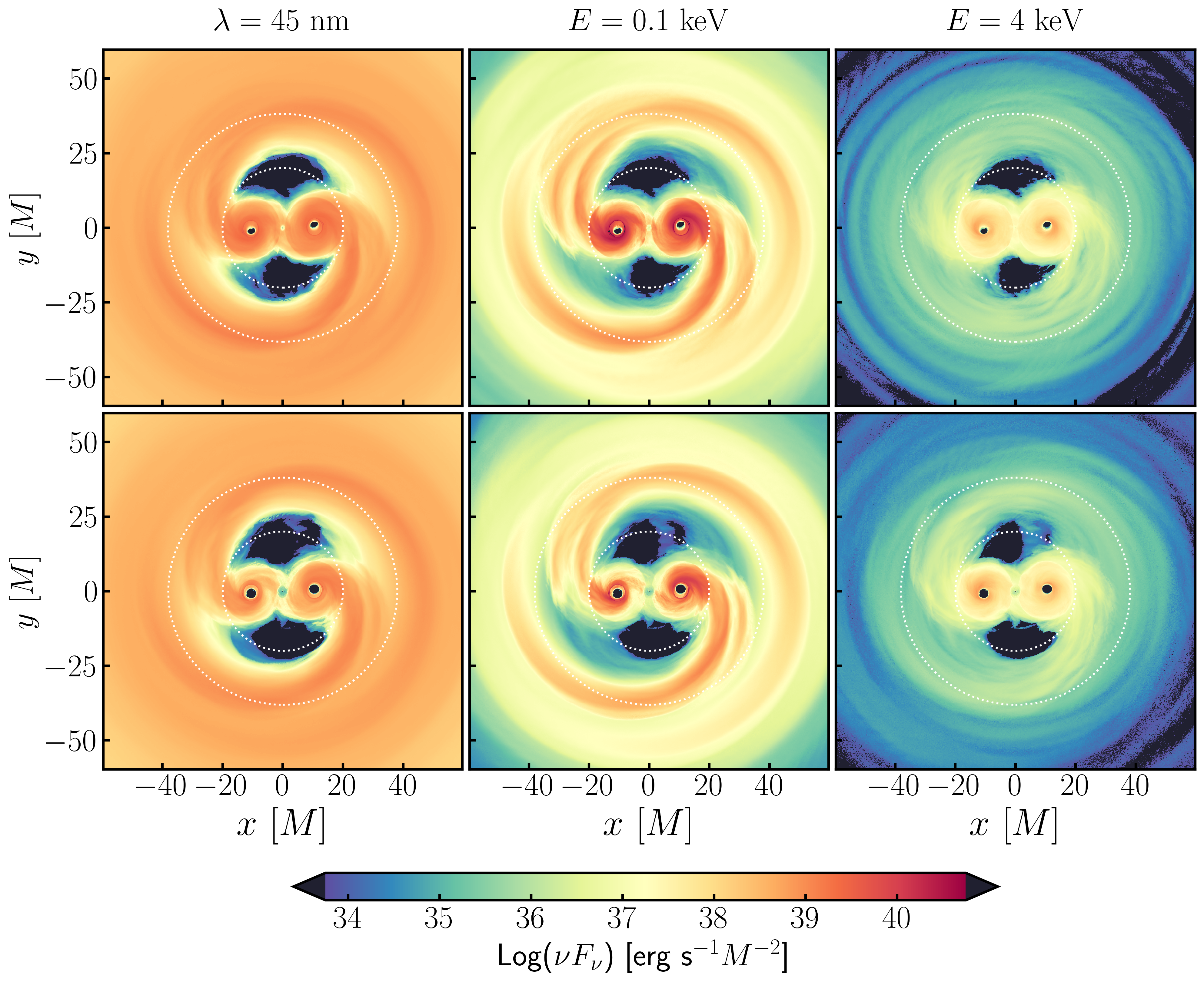}
    \caption{Surface brightness map for the inner $\sim 50~M$ of the accretion flow for three different frequencies, averaged during the fifth orbit. The inner and outer white circumferences represent the outer boundary of the mini-disk zone and the inner boundary of the CBD, respectively. Upper panel: \Ss simulation. Lower panel: \Ns simulation. The total mass of the system is $10^6M_\odot$.}
    \label{fig:img_orbit5}
\end{figure*}

\begin{figure}
    \centering
    \includegraphics[width=0.99\linewidth]{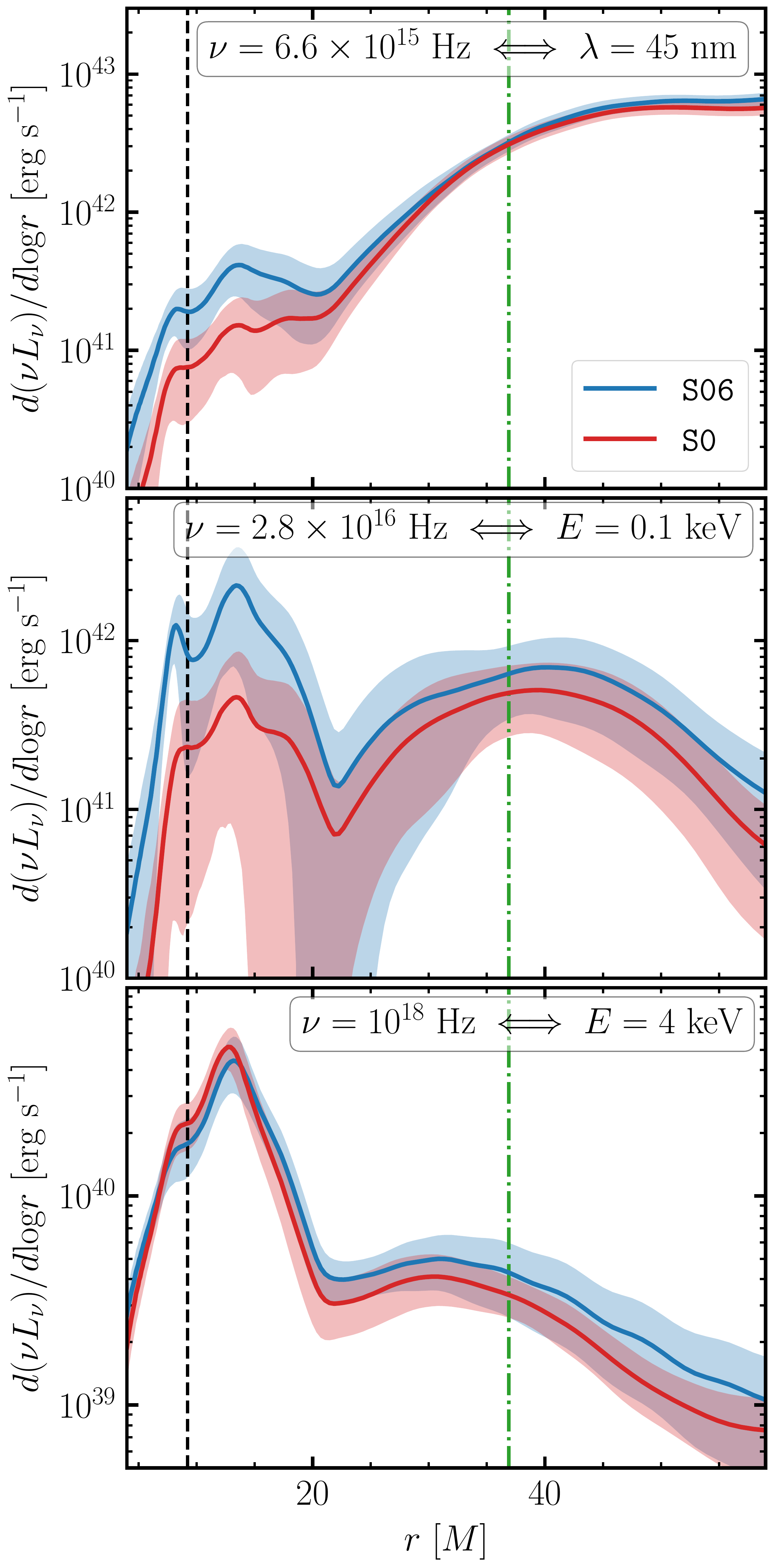}
    \caption{Luminosity per unit logarithmic interval of radius averaged in time. The average for \Ss (\Ns) was done from $t=2000M$ until the end of the \Ns simulation: $t=6430M$. The three panels show the three different frequencies displayed in Figure~\ref{fig:img_orbit5}. The shadowed areas represent one standard deviation from the mean value. The black dashed and green dotted--dashed vertical lines indicate the mean position of the black holes ($r=r_{12}/2$) and the inner radius of the CBD ($r=2r_{12}$), respectively.}
    \label{fig:dLdr}
\end{figure}

To compare the emission of \Ss and \Ns simulations, we first show in Figure~\ref{fig:img_orbit5} brightness maps at three different frequencies, time-averaged over the fifth binary orbit and displayed in the binary's co-rotating frame. The choice of the fifth orbit is arbitrary, but it secures that both simulations have passed the initial transient.
The upper panel corresponds to \Ss and the lower panel to \Ns. Both simulations share several emission characteristics.
The UV map (first panel, $\nu=6.6\times 10^{15}~\hz$ or $\lambda = 22.0~{\rm nm}$) shows a surface brightness that is nearly constant from the CBD down to the mini-disks, except for the low-density cavity. As a result, the CBD dominates the total flux due to its much larger area (see Figure~\ref{fig:spectrum}). The far-UV emission (second panel, $\nu=2.8\times 10^{16}~\hz$ or $\lambda = 99.3~ {\rm nm}$) reveals higher temperature regions, and mini-disks and streams now dominate. The thermal emission from the CBD is lower, indicating that this frequency has passed its thermal maximum. In X-rays (third panel, $\nu=10^{18}~\hz$ or $h\nu = 4.1~\rm{keV}$), the emission is from the optically thin regions and is dominated by the mini-disks, showing that a significant fraction of their emission occurs in the corona rather than in the optically thick disk.

The main difference between the \Ss and \Ns panels is in the mini-disks' thermal emission. While the X-ray map is fairly similar in both scenarios, the first two panels show that the thermal emission from the mini-disks and streams is $\sim 3$ times higher in \Ss.
This happens because the smaller value of $l_{\rm crit}$ when the BHs spin channels a larger fraction of the accretion rate into orbits within the mini-disks rather than plunging orbits.  In order to move inward, the matter held within the mini-disks must lose angular momentum, and the processes that transfer angular momentum are accompanied by dissipation whether they are magnetic stresses associated with MHD turbulence or shocks. Furthermore, the streams in \Ss are brighter than in \Ns both where they strike the CBD and as they fall toward the mini-disks.

To investigate the radial distribution of luminosity, we integrate the intensity maps in the azimuthal coordinate of the image. Figure~\ref{fig:dLdr} depicts $d(\nu L_\nu)/d{\rm log} r$ in the radial range $4$--$60M$ for the three frequencies considered in Figure~\ref{fig:img_orbit5} (here $r$ is the radius from the center of mass). The upper panel shows the luminosity in the UV band, where the CBD dominates the emission. The mini-disks and streams in \Ss are brighter than in \Ns, but the difference fades and becomes insignificant in the CBD region. In the far-UV band (middle panel), the mini-disks surrounding the spinning black holes are a factor of two to five times brighter than in the nonspinning simulation. The relative variance in the emission is also higher than at lower frequencies, particularly at the boundary between the mini-disk and stream regions. This could be a result of our different prescriptions for the cooling function in the mini-disks and CBD.
On the contrary, the optically thin emission (lower panel) from the mini-disks is nearly indistinguishable in the two cases. The optical depth to Compton scattering in the corona is always unity, and hence the coronal surface density is $1/\kappa_{\rm T}$, where $\kappa_{\rm T}$ is the Thomson opacity. Since the area of the mini-disks changes very little with spin, so does the total mass in the corona. Therefore, the near-constancy of the coronal luminosity with respect to spin suggests that the dissipation rate per unit mass in the corona is also spin-insensitive. For the same reason, the coronal luminosity should also not be overly affected by a change in accretion rate provided the mini-disks' surface density remains $>1/\kappa_{\rm T}$ over most of their area.

\subsection{Time Dependence of the Emission}

\begin{figure*}
    \centering
    \includegraphics[width=0.49\linewidth]{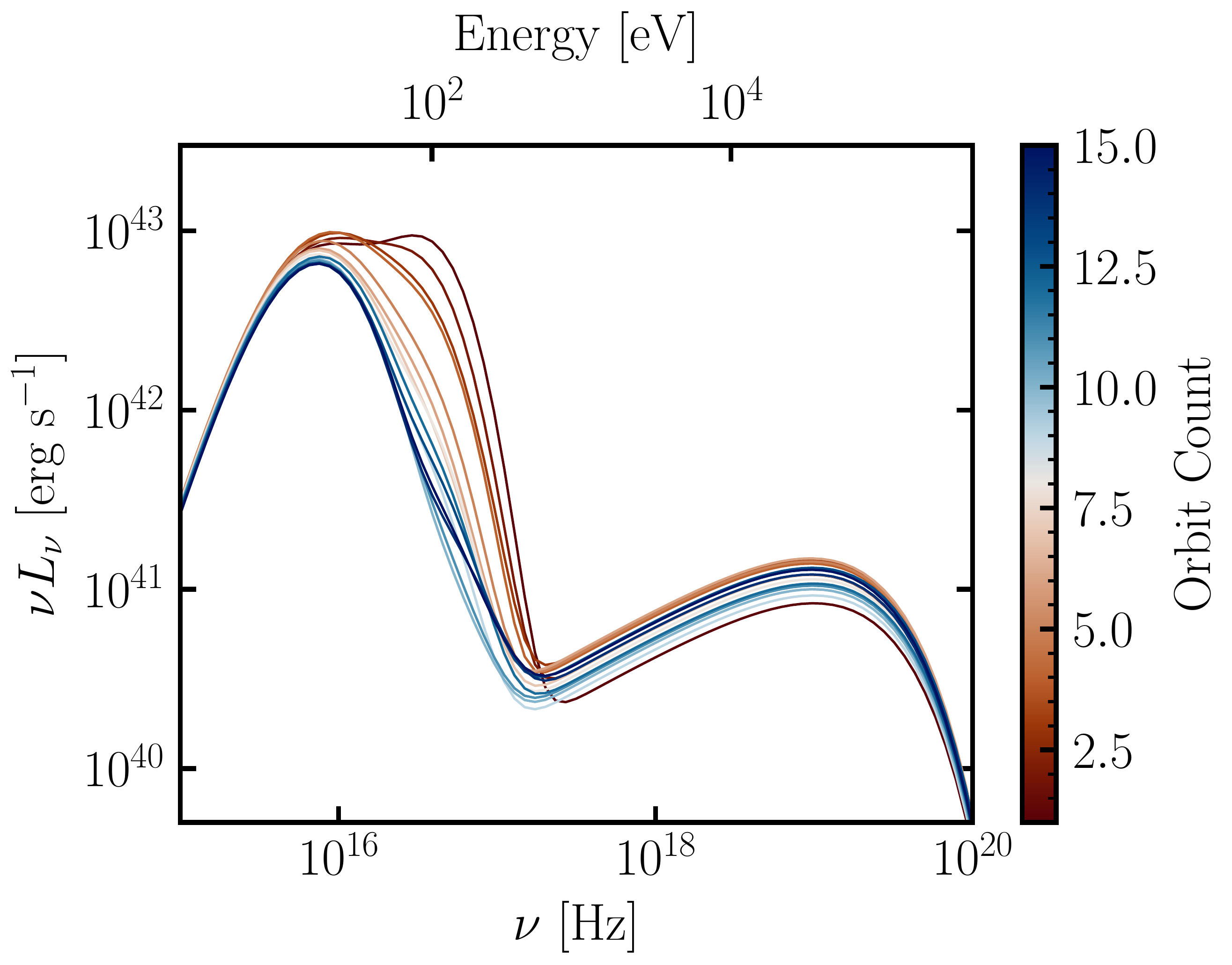}
    \includegraphics[width=0.49\linewidth]{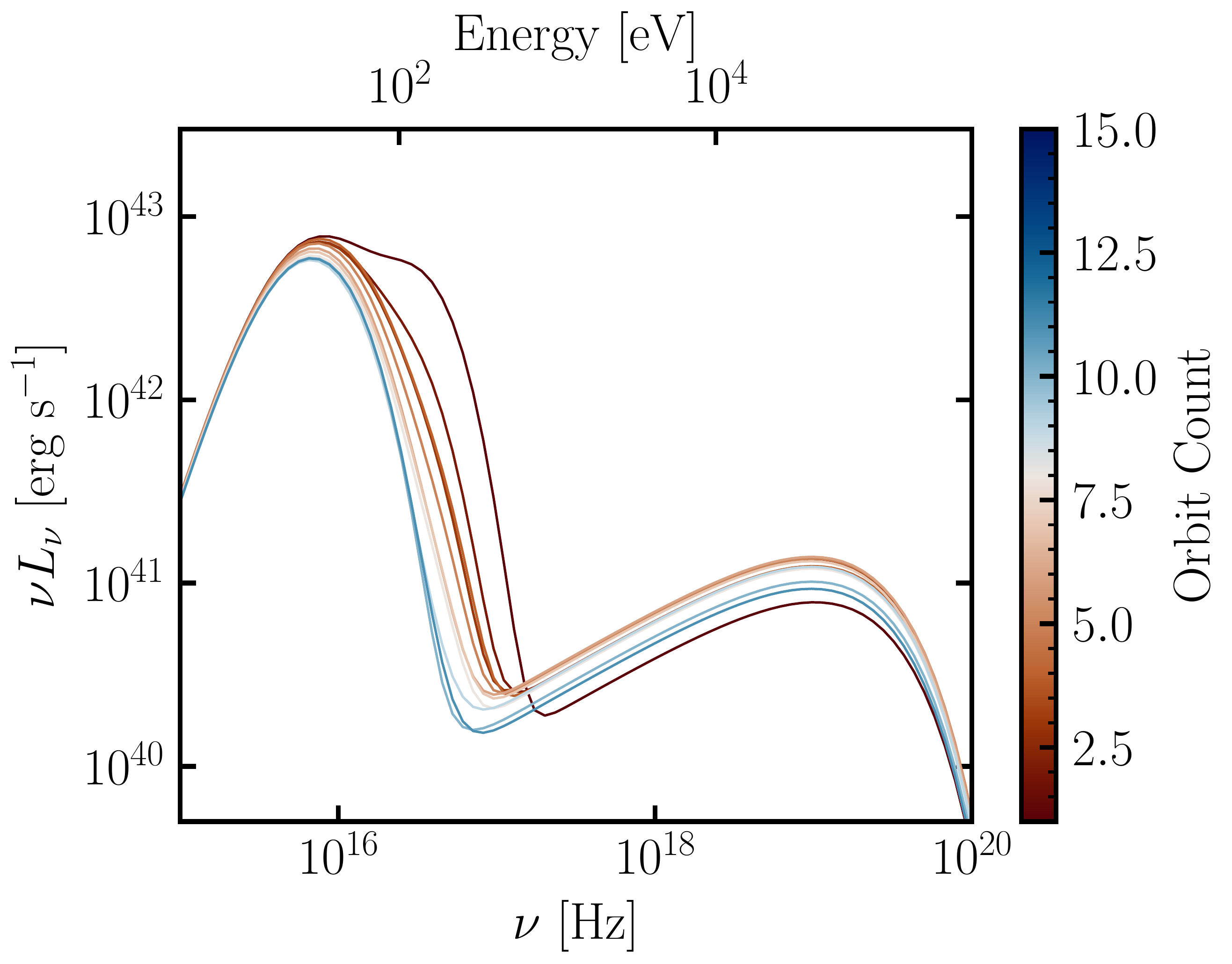}
    \caption{Time-averaged SEDs for each completed orbits. Left panel: \textsc{S06}. Right panel: \textsc{S0}.}
    \label{fig:spectrum_averaged}
\end{figure*}

Due to the complex dynamics associated with orbital motion, accretion flows onto SMBBHs are intrinsically variable. It is worthwhile to investigate how much of this variability is inherited by the EM emission.
In Figure~\ref{fig:spectrum_averaged}, we show the time-average of the spectral energy distributions (SEDs) over each binary orbit. The first orbits correspond to the transient phase, during which the mini-disk emission unphysically increases rapidly and then gradually decreases and stabilizes, becoming quasi-steady after the $\sim 4$-th orbit. After the $\sim 8$-th orbit, the mini-disks in the \Ns simulation barely contribute to the thermal spectrum. In \Ss, on the other hand, their contribution in the far-UV band is nonnegligible until the last ($15$th) orbit, owing to the persistence of a disk-like structure \citep{Combi2021b}.

\begin{figure}[h!]
    \centering
    \includegraphics[width=0.99\linewidth]{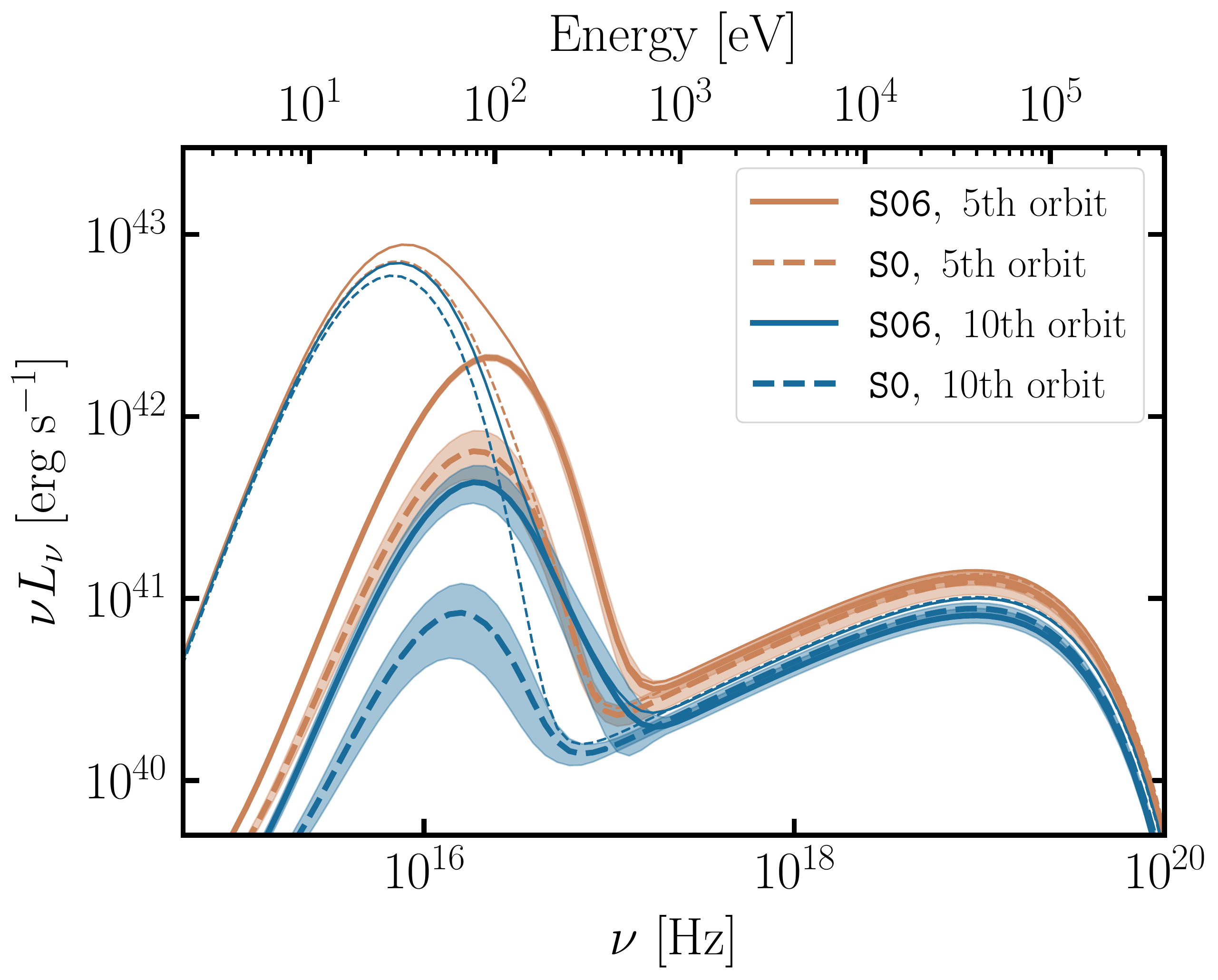}
    \caption{Time-averaged SEDs during the fifth and $10$th orbit for both simulations. The thick lines correspond to the mini-disk spectrum, and the thin lines correspond to the total one. Shaded regions following the curves associated with the mini-disks cover one standard deviation of variation during the orbit.}
    \label{fig:spectrum_comparison}
\end{figure}

In Figure~\ref{fig:spectrum_comparison}, we compare the SEDs of the two simulations for the averaged fifth and $10$th orbits. Though the bolometric luminosity, which is dominated by CBD emission, is comparable for both simulations, the mini-disk's thermal peak (shown in thicker lines) is $\sim 3-5$ times brighter for the spinning black holes. Once again, this difference is explained by the fact that, in the spinning case, the Keplerian angular momentum at the ISCO is smaller, so that a greater amount of mass must lose more angular momentum, dissipating more energy in the process.

\begin{figure*}
    \centering
    \includegraphics[width=0.98\linewidth]{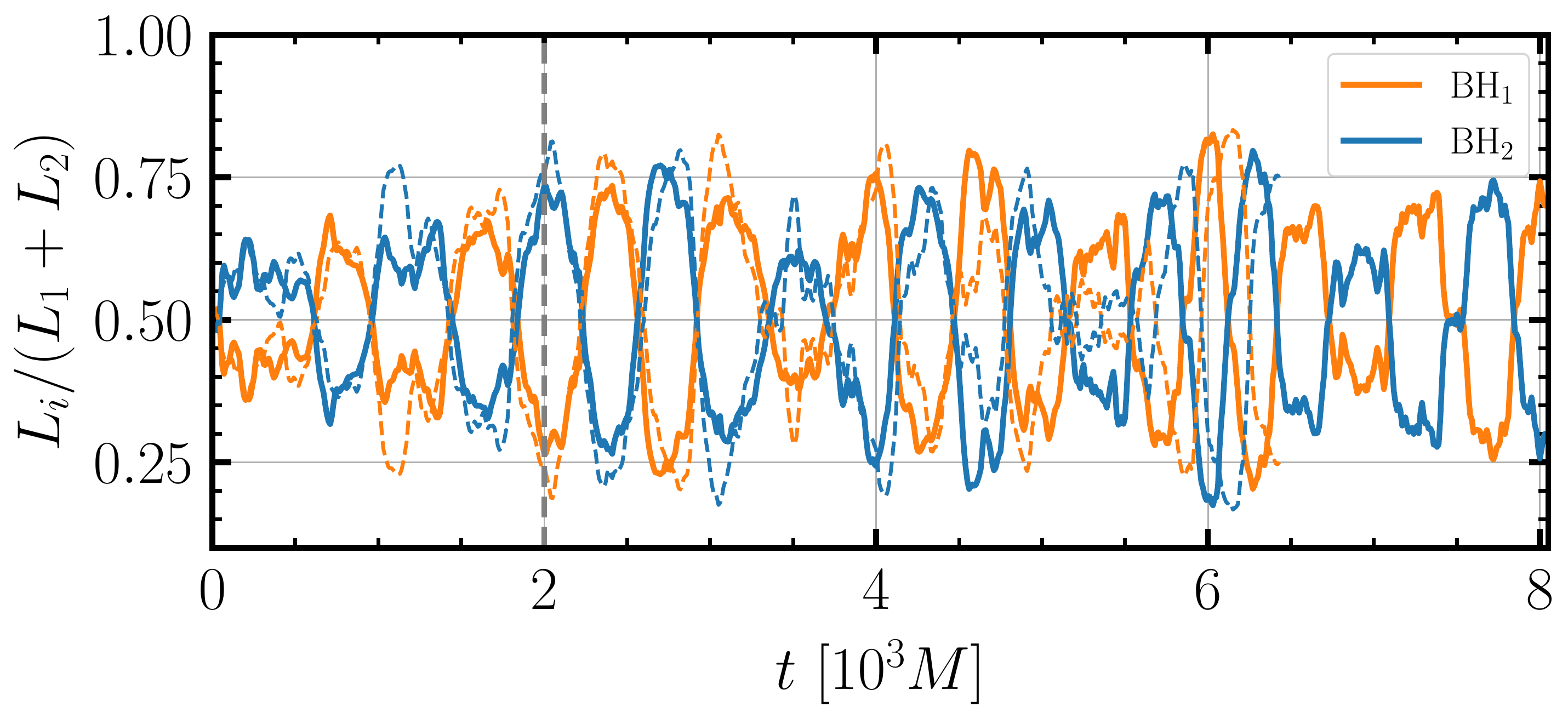}
    \caption{Normalized luminosity of the mini-disks for both simulations. The dashed lines correspond to \Ns and solid lines to \Ss. The gray dashed vertical line shows the end of the transient phase.}
    \label{fig:lc_mds}
\end{figure*}

\begin{figure}
    \centering
    \includegraphics[width=0.99\linewidth]{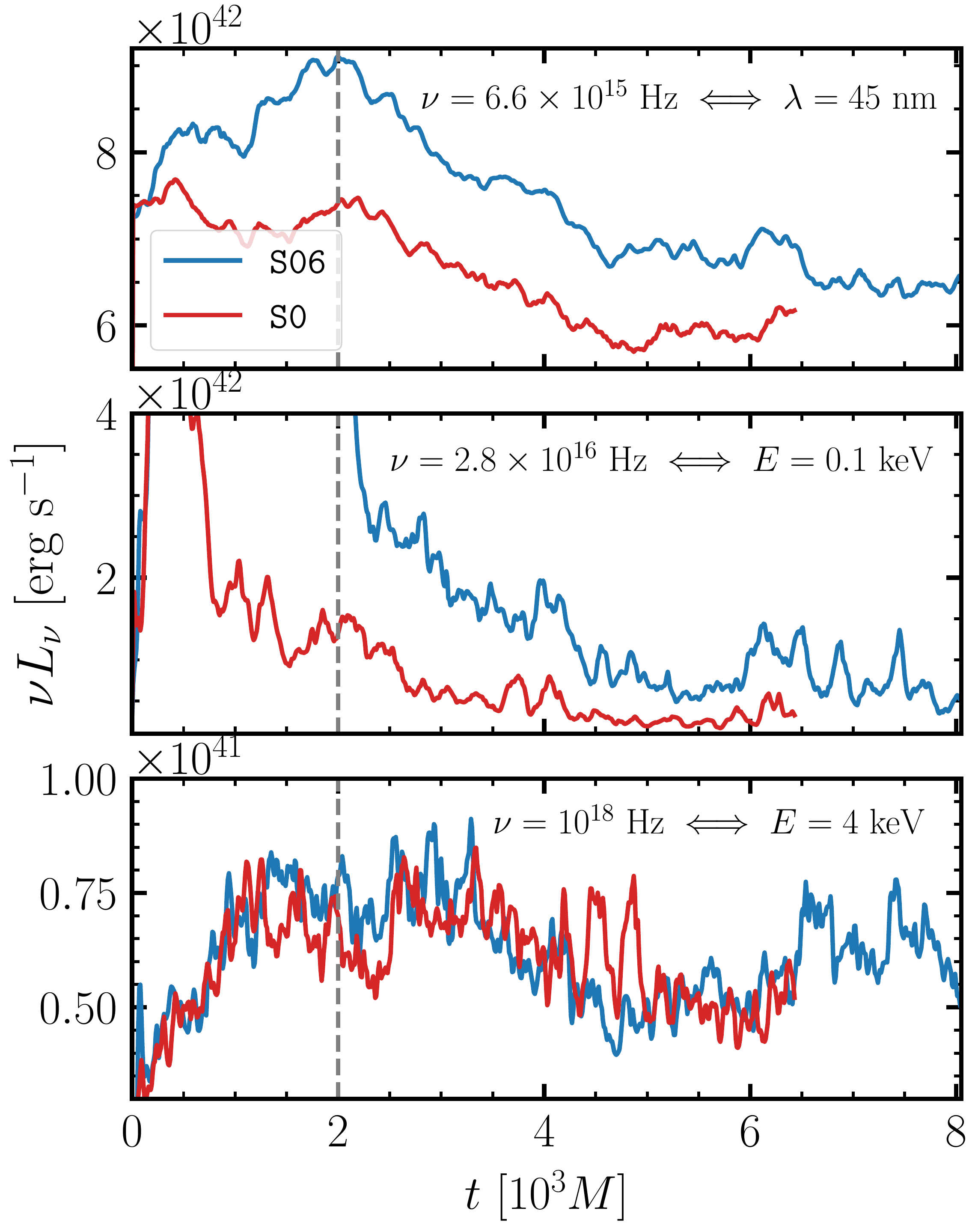}
    \caption{Luminosity as a function of time for \Ss at three different frequencies: UV (upper panel, $\nu = 6.5 \times 10^{15}$ Hz), far-UV (middle panel, $\nu = 2.9 \times 10^{16}$ Hz), and soft X-rays (lower panel, $\nu=10^{18}$ Hz). The gray dashed vertical lines show the end of the transient phase. Note the different dynamic ranges in each panel: $\sim 50\%$ in the top panel, a multiplicative factor $\sim 8$ in the middle panel, and a multiplicative factor of $\sim 2$ in the bottom panel.}
    \label{fig:lcs}
\end{figure}

\begin{figure}
    \centering
    \includegraphics[width=0.99\linewidth]{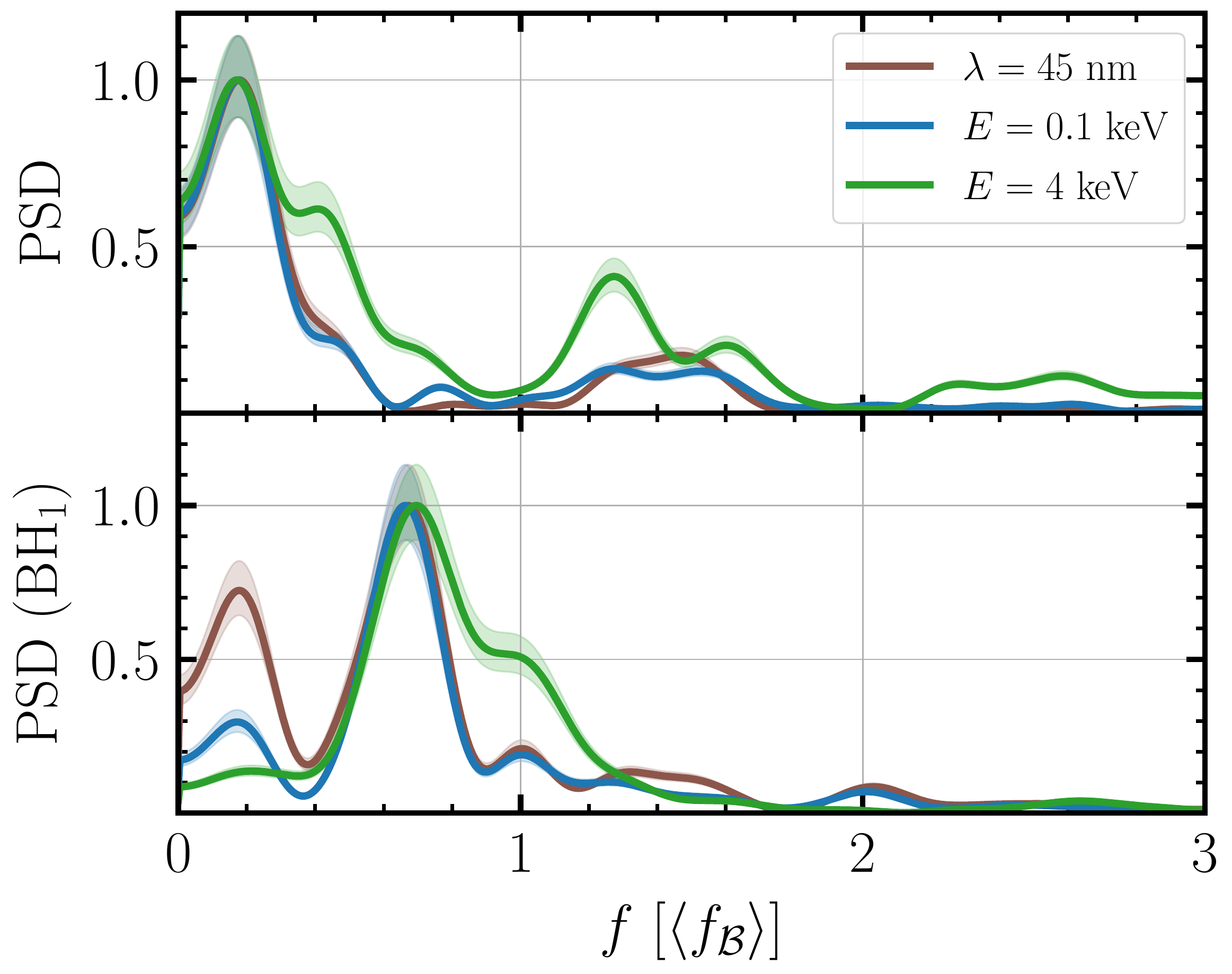}
    \caption{Power spectral density (PSD) of the light curves for \Ss at the three frequencies indicated using a Welch algorithm with Hamming window size of $10M$. The confidence intervals at $3\sigma$ are shown as shadowed areas. The upper panel corresponds to the total luminosity whereas the lower panel takes into account only the emission coming from one of the mini-disks. The mean orbital frequency is $\langle f_\mathcal{B} \rangle=1/505M$.}
    \label{fig:psd}
\end{figure}

\begin{figure}
    \centering
    \includegraphics[width=0.99\linewidth]{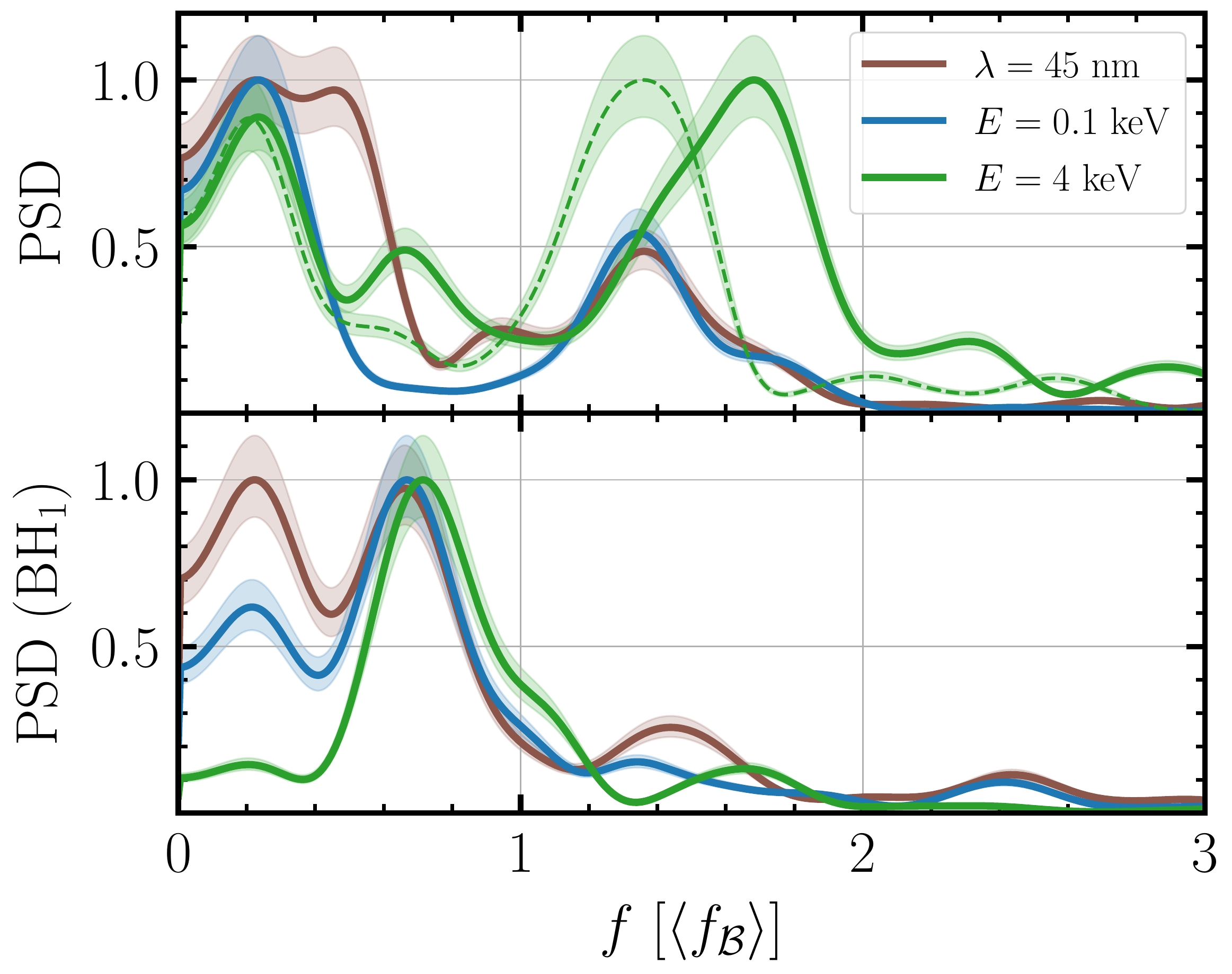}
    \caption{The same plot as in Figure \ref{fig:psd} but for \Ns. The dashed curve in the upper panel corresponds to the light curve at $E=4~{\rm keV}$ with the time corrected for the decreasing period of the system. The mean orbital frequency is $\langle f_\mathcal{B} \rangle=1/530M$.}
    \label{fig:psd_ns}
\end{figure}

We then use the spectral flux at each time to calculate light curves for \Ss and \Ns simulations. The light curves at different frequencies illustrate the variability in the properties of the various contributors to the emission.
The lump modulates the accretion rate and mass in the mini-disks, which is inherited by the luminosity (see Figure~\ref{fig:lc_mds}). Nonetheless, because we would, in principle, observe the joint emission from the two mini-disks, we also look at the variability in the total emission from the system. Figure~\ref{fig:lcs} displays light curves in the UV band (upper panel), where the CBD emission peaks, far UV (middle panel), where the mini-disk emission peaks, and soft X-rays (lower panel), where the optically thin coronal emission dominates (see also Figure~\ref{fig:img_orbit5}).

To further investigate the putative periodicities, we calculate the power spectral density (PSD) of the three light curves for each simulation, using a time sampling of $10M$. The spinning case is depicted in the upper panel of Figure~\ref{fig:psd}, where the PSD has been normalized to its maximum value, and the frequency is shown in units of the mean orbital frequency. The mean period for \Ss is $\langle P \rangle \approx 505M$, and for \Ns it is $\langle P \rangle \approx 530M$. The three curves peak at $f\sim 0.2f_\mathcal{B}$, which approximately corresponds to the frequency of the lump's radial oscillation, since the lump follows a slightly eccentric path around the cavity \citep{lopezarmengol2021, noble2021}; at the point of closest approach, the rate of matter falling into the cavity increases and so does the mini-disk luminosity. A second, smaller peak at $f \sim 1.4f_\mathcal{B}$ is visible primarily in X-rays. This is twice the beat frequency and corresponds to the accretion event that happens when one of the two mini-disks passes near the lump. In the lower panel of Figure \ref{fig:psd}, we show the PSD for one mini-disk alone. The peak at the beat frequency is quite visible here, and it is even greater than the one associated with the lump's radial oscillation.

In Figure \ref{fig:psd_ns}, we show the PSD for \Ns. The three curves in the upper panel, which represent the total emission at different frequencies, have some differences with respect to the spinning case. The low-frequency variability is less predominant since \Ns runs for a shorter period of time, not long enough to resolve this periodicity well. The peak at $\sim 2 f_{\rm beat}$, on the other hand, is more intense than in \Ss because the mini-disks are less massive and deplete completely during a beat period. Another distinct feature of this peak is that it is shifted to higher frequencies in the X-ray curve. This appears to be an effect of the binary frequency increasing during the inspiral. If we account for this time-dependent orbital frequency in the light-curve data by redefining the unit of time to be the instantaneous binary orbital frequency, 
 the peak in the PSD shifts back to $f \sim 2 f_{\rm beat}$ (see the green dashed curve in the upper panel of Figure \ref{fig:psd_ns}).
 This periodic signal is likely stronger at later times when the orbital period is shorter than $\langle P \rangle$. The lower panel, which depicts the PSD for a single mini-disk, shows no significant differences from the spinning case.

\section{Discussion} \label{sec:discussion}

We have described in Sec. \ref{sec:results} the primary characteristics of the emitted light in our simulations by producing brightness maps, spectra, and light curves. We have also analyzed the distinctions between spinning and nonspinning cases. A key question to address is whether there are discernible spectral or temporal signatures that could be used to distinguish SMBBHs from single AGNs.
In this work, we investigate a limited subset of the parameter space: SMBBHs with equal masses of $0.5\times 10^6M_\odot$ (but see Sec. \ref{subsec:observations}) at separations of the order of $\sim 20M$, and for a period of $10-15$ orbits. Additionally, we assumed a moderately high accretion rate and made some simplified, though reasonable, assumptions about how the plasma cools and radiates. Nonetheless, our calculations are the most realistic made to date for this type of scenario, and it is worthwhile to investigate how much our analysis contributes to answering this question.

\subsection{Comparison with Single-disk Analytical models} \label{subsec:mini_disks}

\begin{figure*}
    \centering
    \includegraphics[width=0.9999\linewidth]{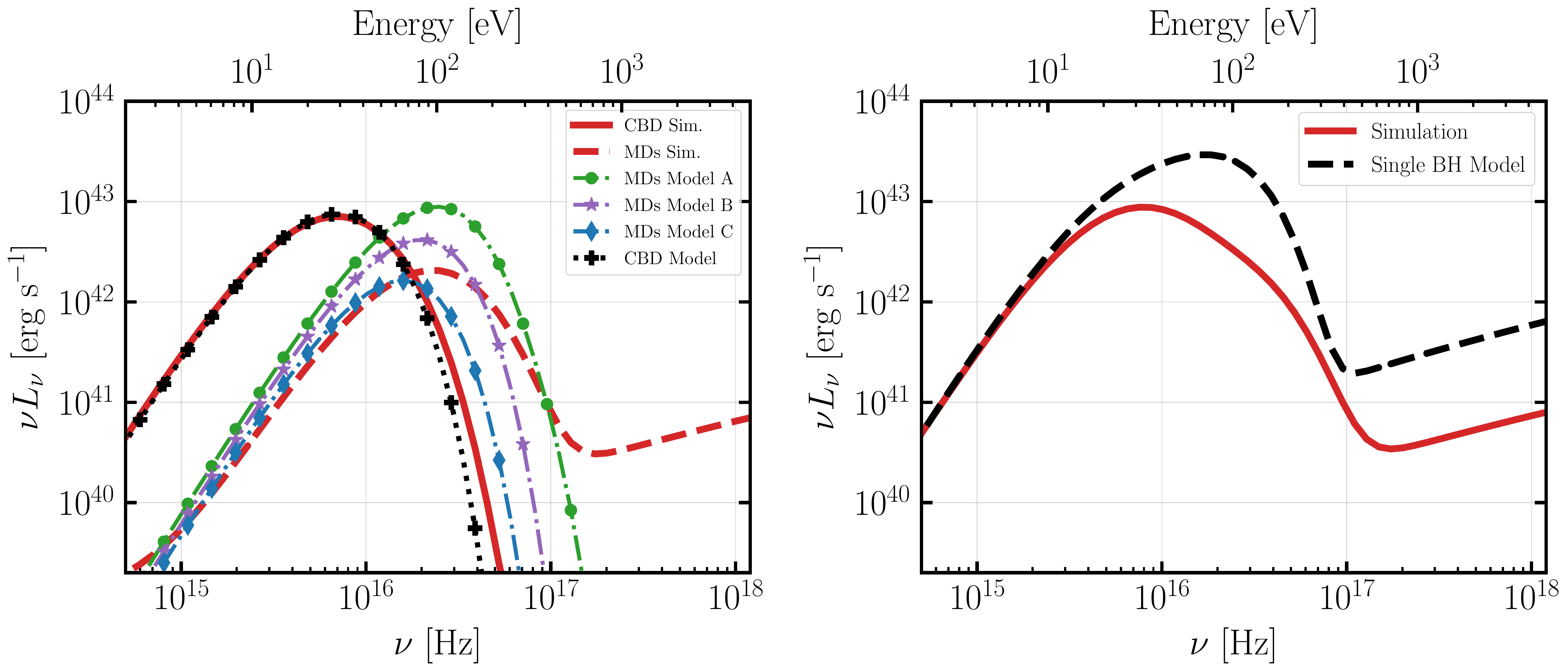}
    \caption{Comparison between the time-averaged SED during the fifth orbit of \Ss simulation and single black hole disk models. Left panel: Comparison of the numerical spectrum of the CBD (solid red line) and the mini-disks (dashed red line) with the following Novikov--Thorne (NT) disk models. The curve denoted as ``CBD Model'' shows the spectrum of an NT disk for a black hole mass of $10^6M_\odot$ and normalized spin $\chi=0.6$, an accretion rate of $0.25\dot{M}_{\rm Edd}$, an inner radius at $2\langle r_{12} \rangle \sim 38M$, and an outer radius at $150M$. Each of the curves denoted as MD ``Model A'', ``B'', and ``C'' show the added spectrum of two NT disks onto equal black holes with a mass of $0.5 \times 10^6M_\odot$ and a normalized spin of $\chi =0.6$. The NT disks have the inner radius at the individual ISCOs, $\sim 3.8m_{1,2}=1.9M$, the outer radius at $0.4\langle r_{12} \rangle \sim 7.6M=15.2m_{1,2}$, and an accretion rate of (Model A: $\dot{m}_1=\dot{m}_2=0.125 \dot{M}_{\rm Edd}$; Model B: $\dot{m}_1=6\times 10^{-2}\dot{M}_{\rm Edd},~\dot{m}_2=6.8\times 10^{-2} \dot{M}_{\rm Edd}$; Model C: $\dot{m}_1=2.3\times 10^{-2}\dot{M}_{\rm Edd},~\dot{m}_2=2.7\times 10^{-2} \dot{M}_{\rm Edd}$). These values correspond to the accretion rate measured in the CBD region, and the total and circularized accretion rates measured onto each black hole, respectively. Right panel: comparison between the total SED derived from the simulation and one for a single black hole of mass $M=10^6M_\odot$ in which we use \citet{Schnittman2016}'s radial luminosity profile. The disk has an accretion rate of $0.25\dot{M}_{\rm Edd}$, an inner radius at $\sim 1.2 r_{\rm H} \sim 2.16 M$, and an outer radius at $150M$. Ten percent of the luminosity arises from an optically thin corona.}
    \label{fig:sim_vs_analytical}
\end{figure*}

Single supermassive black holes accreting at moderately high rates are thought to consist of a geometrically thin, optically thick, and radiatively efficient disk that emits a multitemperature blackbody spectrum peaking in the UV band, and an optically thin hot corona that emits a power-law spectrum at higher energies. The simplest and most used model for the thermal component is the Novikov--Thorne (NT) disk model \citep{NovikovThorne73}, defined by the assumptions of nearly-circular orbits, time-steadiness, azimuthal symmetry, local radiation of dissipated energy, and no stress inside the ISCO.

To determine whether NT models provide a good approximation to the emission from an accreting binary black hole system approaching merger, we compare the ray-traced SED from simulation \Ss, averaged during the fifth orbit, to the face-on SED of various NT models and show the results in the left panel of Figure~\ref{fig:sim_vs_analytical}. We calculate the NT flux as
\begin{equation}
F_{\nu_{\rm obs}} = \frac{2\pi}{d^2} \int_{R_{\rm in}}^{R_{\rm out}} dr ~ r ~g^3(r) ~ B_{\nu_{\rm em}}(T_{\rm eff}(r)),
\label{eq:F_nu_Kerr}
\end{equation}
where $\nu_{\rm obs} = g \nu_{\rm em}$ and 
\begin{equation}
    g:= \frac{(k_\mu)_{\rm obs} u^\mu_{\rm obs}}{ (k_\nu)_{\rm em} u_{\rm em}^\nu} = \sqrt{ | -g_{tt} - 2\Omega g_{t\phi} - \Omega^2 g_{\phi \phi} |},
\end{equation}    
is the redshift factor for face-on emission. Here, $g_{\mu \nu}$ are the components of the Kerr metric in Boyer--Lindquist coordinates and $\Omega$ is the orbital velocity of a Keplerian circular orbit in the Kerr spacetime. Eq. \ref{eq:F_nu_Kerr} neglects light bending, but this phenomenon has little effect on the spectrum for face-on emission.

SMBBH emission comes from both the CBD and the mini-disks.
Not surprisingly, given the CBD's state of quasi-inflow equilibrium and its distance from the nominal ISCO (corresponding to the total mass of the binary), its emission averaged over the fifth orbit is well reproduced by an NT disk extending from $R_{\rm in, circ}=2\langle r_{12}\rangle \sim 38 M$ to $R_{\rm out, circ}=150M$, accreting at a rate of $0.25 \dot{M}_{\rm Edd}$, which are the same values used in our simulations. Quantities enclosed in brackets are averaged during the fifth orbit.  Even though we integrate from the inner edge to the outer part of the disk, for this NT disk model, we have set the stress to zero at $R_{\rm ISCO, circ}=6M$, which corresponds to the system's fictitious ISCO.

To analyze the SED of the mini-disks, we compare them to three different NT models with varying accretion rates. None is a good match to the spectrum we calculate. In all cases, we set the outer radius equal to the truncation radius, $R_{\rm out, md}\sim 0.4\langle r_{12}\rangle \sim 8M=16m_i$, the inner radius equal to the individual ISCOs, $R_{\rm ISCO,md} (a=0.6) = 3.8m_i = 1.9M$, and the mass and spin equal to those of the black holes in the simulation: $(m_i=0.5\times 10^6 M_\odot$, $\chi_i=0.6)$.
At a fixed accretion rate in Eddington units, the frequencies of features are $\propto M^{-1/4}$.

In Figure~\ref{fig:sim_vs_analytical}, MD Model A (red curve) represents the case in which the accretion rate onto both mini-disks is equal to that in the CBD region, $0.25\dot{M}_{\rm Edd}$. For simplicity, we assume that it divides evenly between the two mini-disks\footnote{In the real scenario, however, one mini-disk is typically brighter than the other at any given time; this effect is periodic (see Figure~\ref{fig:lc_mds}) and the variation during an orbit likely averages out this difference.}. The spectrum obtained is $\sim 3.7$ times brighter than the one obtained from \Ss. One source of this large discrepancy is a breakdown in the NT model assumption of inflow equilibrium: the accretion rate in the mini-disks is approximately half that in the CBD region. More precisely, the averaged accretion rates onto the black holes during the fifth orbit are $\sim 6.8 \times 10^{-2} \dot{M}_{\rm Edd}$ and $6 \times 10^{-2}\dot{M}_{\rm Edd}$ when the CBD accretion rate is $0.25 \dot{M}_{\rm Edd}$.

However, this accretion rate contrast does not completely explain the shortfall. Model B shows the combined spectrum of two NT mini-disk models with the actual accretion rates. They are still $\sim 1.7$ times brighter than the numerical spectrum, indicating that the mini-disks have a lower radiative efficiency than the NT disk. 
At least part of this diminished radiative efficiency is due to some of the accreting matter at each radius having less angular momentum than the value required for a circular orbit at that radius, i.e., $l(r) < l_{\rm K}(r)$. This material, which follows a decidedly noncircular orbit, is able to reach the event horizon with higher orbital energy (lower binding energy) than matter following stable circular orbits.
To distinguish the luminosity from the fluid that follows quasi-circular orbits from that radiated by the fluid on noncircular orbits, we define the `circularized' accretion rate as the rate delivered by matter with $l(r) \geq l_{\rm K}(r)$, and averaging from $r_{\rm ISCO}$ to $r_{\rm trunc}$. The `circularized' accretion rates are $2.3 \times 10^{-2} \dot{M}_{\rm Edd}$ and $2.6 \times 10^{-2}\dot{M}_{\rm Edd}$ for the two black holes, respectively.

Model C shows the spectrum for two NT disks with the circularized accretion rates of the real mini-disks, but this model still departs from the simulation spectrum in significant ways; its luminosity is a factor $\sim 1.5$ lower than the simulated SED, and it is significantly ``softer'' at frequencies above the peak.

In fact, the three analytical models all produce softer thermal spectra, which could be an effect of the higher temperatures achieved by the shocked plasma in the mini-disks.
Another difference is that in the simulation, part of the cooling occurs above the thermalized photosphere producing a different spectrum: a power law up to hard X-rays. In fact, the absence of hard X-rays with significant luminosity is one of the principal failings of the NT model prediction for ordinary AGNs.

The analysis above shows that a typical mini-disk behaves differently than an NT disk onto a single black hole for the separations considered.
Moreover, analytical models designed for larger separation binaries, in which the mini-disk accretion rate matches the CBD accretion rate and the angular momentum of all material delivered to the mini-disks is large enough that it cannot plunge, predict a clear `notch' in the thermal spectrum (see, e.g., \citealt{Roedig2014}) due to the absence of emission from the region $0.4r_{12}<r<2r_{12}$. In contrast, the low radiative efficiency of the mini-disks at these separations completely overcomes this effect in our simulation. This is likely to change with increasing spins or at larger black hole separations.

It is also worthwhile to compare the predicted emission from an accreting binary black hole system with that of an accreting single black hole system, given the same global parameters: the total mass, outer radius, and accretion rate.
Detailed GRMHD simulations of single black hole accretion disks show that the stresses in the disk do not completely vanish at the ISCO, and the dissipation profile deviates from NT. \citet{Schnittman2016}
derived a very good analytical fit for the radial luminosity profile of single black hole disks in a steady state, using
ray-tracing of GRMHD simulations with cooling similar to that used here (see Equation (3) in the Appendix of that paper). They found that 
$\sim 10\%$ of the emission arises from an optically thin corona rather than from the optically thick disk. Based on these results, we show in the right panel of Figure \ref{fig:sim_vs_analytical} a comparison between the SED from our simulation data and the SED from a single black hole disk model that mimics the results from the GRMHD+postprocessing simulations by \citet{Schnittman2016}. Here, we use their analytical fitting function for the radial luminosity profile and assume that $10\%$ of the luminosity is in the form of a power law with an exponential cutoff, with the same shape as in our simulation. The mass of the black hole is $M=10^6M_\odot$, the accretion rate is $0.25\dot{M}_{\rm Edd}$, the inner radius is $R_{\rm in}=1.2 r_{\rm H} \approx 2.16 M$, and the outer radius is $R_{\rm out}=150M$.

The total SED from the binary looks significantly different from that of the single black hole: the latter has a single broad peak whose luminosity is $\sim 3$ times greater than the binary's. Moreover, the binary's spectrum peaks at a frequency $\sim 3.4$ times smaller than the single black hole's spectrum and has a different slope between the frequency of the maximum and the frequency where the corona starts to dominate. In this region, the binary's SED is a broken power law with a break at the mini-disks' peak. Above this frequency, the spectrum softens but not so abruptly as in the case of a single black hole disk.

The coronal luminosity is also lower for the binary because the bulk of the coronal emission comes from the mini-disks, where the accretion rate is lower than in the CBD and the overall radiative efficiency is lower than that of the single black hole disk (See Sec. \ref{subsec:eff}). These two facts also translate into the fact that the ratio of the luminosity from the inner region ($r \lesssim 20M$) to the total luminosity (up to $r=150M$) is $\lesssim 0.2 $ for the binary, whereas it is $\gtrsim 0.5$ for a single black hole accretion disk.

\subsection{Radiative Efficiency} \label{subsec:eff}

\begin{figure}
    \centering
    \includegraphics[width=0.9999\linewidth]{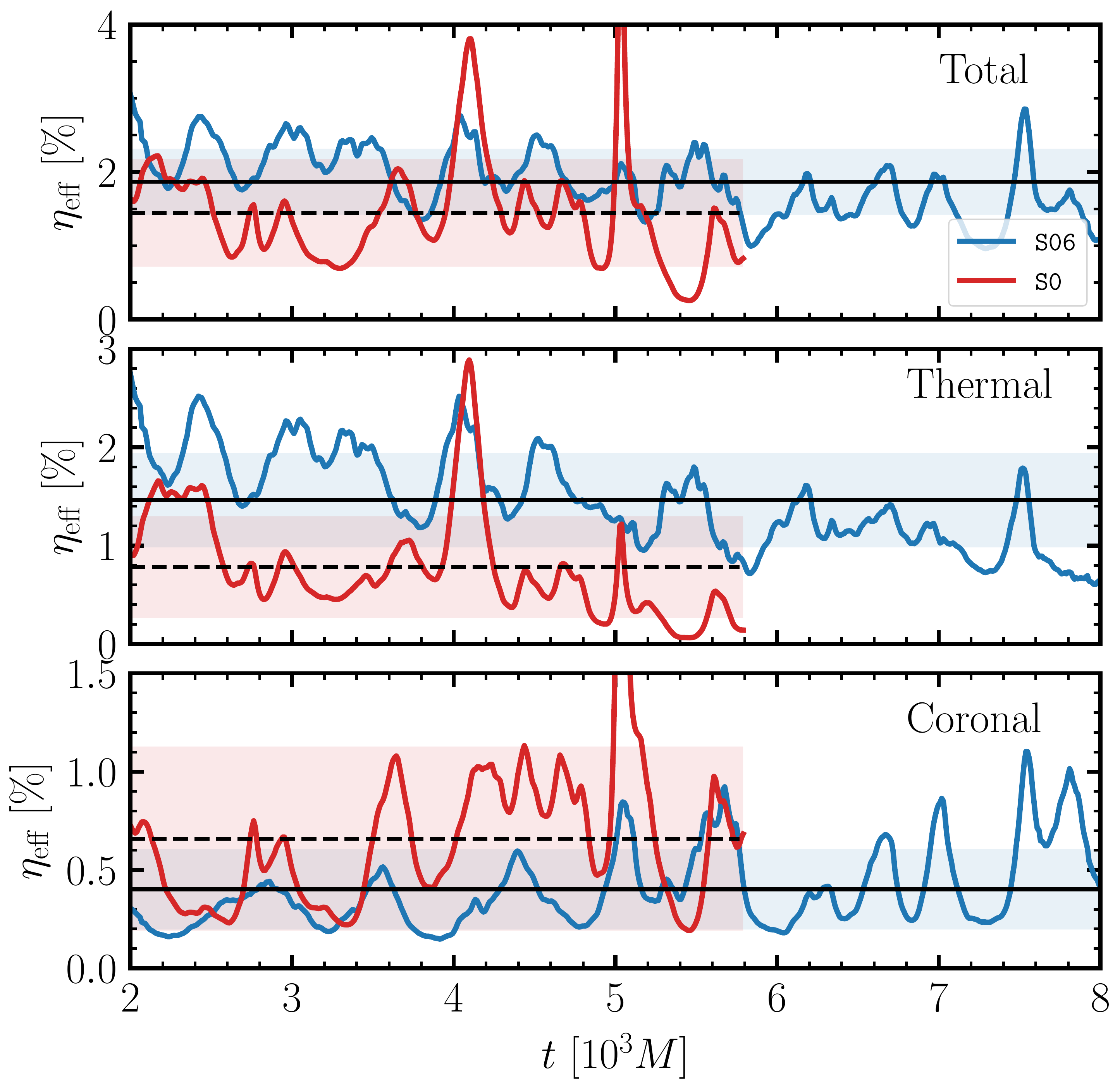}
    \caption{Mini-disk radiative efficiency, defined as $\eta_{\rm eff}\equiv L/\dot{M}c^2$, for both simulations. Upper panel: total efficiency. Middle panel: efficiency of the thermal emission. Lower panel: efficiency of the coronal emission. The horizontal red line and shadow area show the mean value and the standard deviation.}
    \label{fig:eff_disk}
\end{figure}

We have assumed an accretion rate of $0.25 \dot{M}_{\rm Edd}$ in the CBD, which would correspond to a luminosity of $0.25 L_{\rm Edd}$ for a typical AGN with a standard radiative efficiency of $10 \%$. This approximately coincides with the face-on value predicted by the NT model for a radiatively efficient disk onto a black hole with normalized spin $\sim 0.5$.
The bolometric luminosity, which is dominated by the CBD emission, is $\approx 0.1 L_{\rm Edd}$ in our simulations. Thus, the global radiative efficiency is $\approx 4\%$. According to the results discussed in Sec. \ref{subsec:mini_disks}, this contrast is the result of a combination of two effects: the lower accretion rate in the cavity and the mini-disks' low radiative efficiency.

To examine the mini-disks' radiative efficiency in greater detail, we choose one black hole, ${\rm BH}_1$, and calculate the accretion rate onto it in CU as
\begin{equation}
    \dot{M}_{{\rm BH}_1}(t) = \oint_{r_{\rm H}} d\bar{A} \, u^{\bar{r}} \rho,
\end{equation}
where the overbars denote harmonic coordinates centered at the hole. We transform the accretion rate to physical units and calculate the accretion power as a function of time: $L_{{\rm acc, BH}_1}(t) \equiv \dot{M}_{{\rm BH}_1}(t) c^2$. Then, we calculate the radiative efficiency of the mini-disk as $\eta_{\rm eff} \equiv L_{{\rm MD}_1}/L_{{\rm acc, BH}_1}$.

In Figure~\ref{fig:eff_disk}, we show the mini-disk radiative efficiency as a function of time for both simulations. The upper, middle, and lower panels depict the total ($\nu \in (0,\infty )$ Hz), thermal ($\nu \in (0, 2\times 10^{17})~{\rm Hz}$), and coronal ($\nu \in (2\times 10^{17}, \infty)$ Hz) efficiencies, respectively. The three plots show a quasiperiodic oscillatory behavior of the efficiency, with the effect being more pronounced for the \Ss simulation. The spin of the black holes has little effect on the CBD in terms of modifying the bolometric emission, but it does have an effect on the emission from the mini-disk. The mean values for the total, thermal, and coronal radiative efficiency of the mini-disks are $\approx 1.9\%$ ($1.4\%$), $\approx 1.5 \%$ ($0.8\%$), and $\approx 0.4\%$ ($0.7\%$), respectively, for the \Ss (\Ns) simulation. For an NT disk extending from the ISCO to $\sim 0.4 \langle r_{12} \rangle$, where $\langle r_{12} \rangle \approx 18M = 36m_1$, the radiative efficiency calculated in the same way as in our simulation (face-on emission) is $\eta_{\rm NT}(\chi=0.6)\approx 4 \%$ and $\eta_{\rm NT}(\chi = 0)\approx 1.7 \%$ for a spinning and a nonspinning black hole, respectively. The thermal radiative efficiency of the mini-disks is $\sim 40-60 \%$ of that predicted by the NT model for the same spin and disk extension in both \Ss and \Ns. On the other hand, the total radiative efficiency of the mini-disks relative to the NT prediction is a fraction $\sim 55\%$ for \Ss and $\sim 90\%$ for \Ns.

The greater fraction of NT efficiency generated in the \Ns mini-disks is due to the persistence of a coronal region even when the surface density in the mini-disk is relatively small. In other words, for equal accretion rate, more matter is below the thermalized Thomson photosphere in \Ss, resulting in a lower fraction in the corona, and vice versa for \Ns. These results might drastically change during other stages of the inspiral. In particular, at earlier stages and beyond some black hole separation, the single disk's predictions probably would be recovered.

\subsection{Comparison with Previous Works} \label{subsec:otherworks}

Previous attempts to model EM observables from SMBBHs have been based upon 2D-viscous Newtonian simulations  \citep{Farris15a, Farris15b, Tang2018, westernacher2021}, 2D-{inviscid} GRHD simulations of isolated mini-disks in a Kerr spacetime, i.e., without any tidal gravity from the companion \citep{RyanMacFadyen17}, and fully relativistic 3D-MHD simulations \citep{dAscoli2018}.

Our work agrees with the Newtonian 2D-viscous calculations in finding that the shock created when stream material is propelled back out to the CBD can create strong enough heating to influence the output spectrum.  However, we disagree in several respects.
The thermal peaks of the spectra in \citet{Farris15a} and \citet{Tang2018} are shifted to higher frequencies compared to our results (see, e.g., Fig. 2 in \citealt{Tang2018}) because the effective temperature in these works is much higher than our estimates. The reason for this is that they derive the effective temperature assuming the disk pressure is dominated by gas pressure. If radiation pressure is significant, as it likely is for near-Eddington accretion rates and separations out to several hundred gravitational radii, this approach overestimates the effective temperature. In our simulations, the effective temperature is linked directly to the actual dissipation rate (from shocks and turbulence).  Note that \citet{westernacher2021} made efforts to account for this effect, but also assumed a very high accretion rate, $10\dot{M}_{\rm Edd}$.

In some of this work, a spectral notch appears \citep{Tang2018} but not in others \citep{Farris15a}.  When it is absent, it is because radiation from the shocked streams fills the gap between the CBD and mini-disk peaks. In our simulations, although the stream emission does fall in this gap, it is not strong enough to fill it; the notch disappears because the mini-disks are too faint. In the Newtonian calculations, the mini-disks are much more luminous relative to the CBD because they are able to maintain inflow equilibrium and do not permit rapid inflow because their truncation radii are very large compared to their ISCOs.
A final point of contrast in predicted spectra is that these papers considered only thermal emission, whereas we included also coronal X-rays.


There are also points of both agreement and disagreement in the predicted light curves. The equal-mass circular cases considered by \citet{westernacher2021} lack a signal at $2f_{\rm beat}$ because their cavity is too large and offset for a beat frequency to form. Instead, they found two modulations in the light curves: a slow one, associated to the lump orbital frequency, and a fast one with a frequency of $1$--$1.5 f_\mathcal{B}$, whose origin remained undetermined. Although this fast frequency is close to the value of $2f_{\rm beat} \sim 1.4 f_\mathcal{B}$ that we find, theirs seemed to decrease and tend to the orbital frequency as they increased the resolution and shrinked the mini-disk's sink region. They, therefore, argued that the offset from the binary orbital frequency might be either an under-resolved orbital frequency or a phenomenon dependent on other approximations from their approach. Because our results rely on actual accretion dynamics, rather than a sink prescription, the signal at $2f_{\rm beat}$ cannot be an artifact of this kind. Moreover, the same beat frequency features have been found in many previous calculations \citep{Shi2012,Noble12,Bowen2019}.

We extended the calculations made by \citet{dAscoli2018} in two ways: adding a case with spinning black holes to the Schwarzschild case examined in that paper and lengthening the run from three orbits to more than $11$.
The greater duration enabled us to identify truly periodic features in the emission after the system had reached a quasi-steady state.

\subsection{Observational Prospects} \label{subsec:observations}

\begin{figure}
    \centering
    \includegraphics[width=0.99\linewidth]{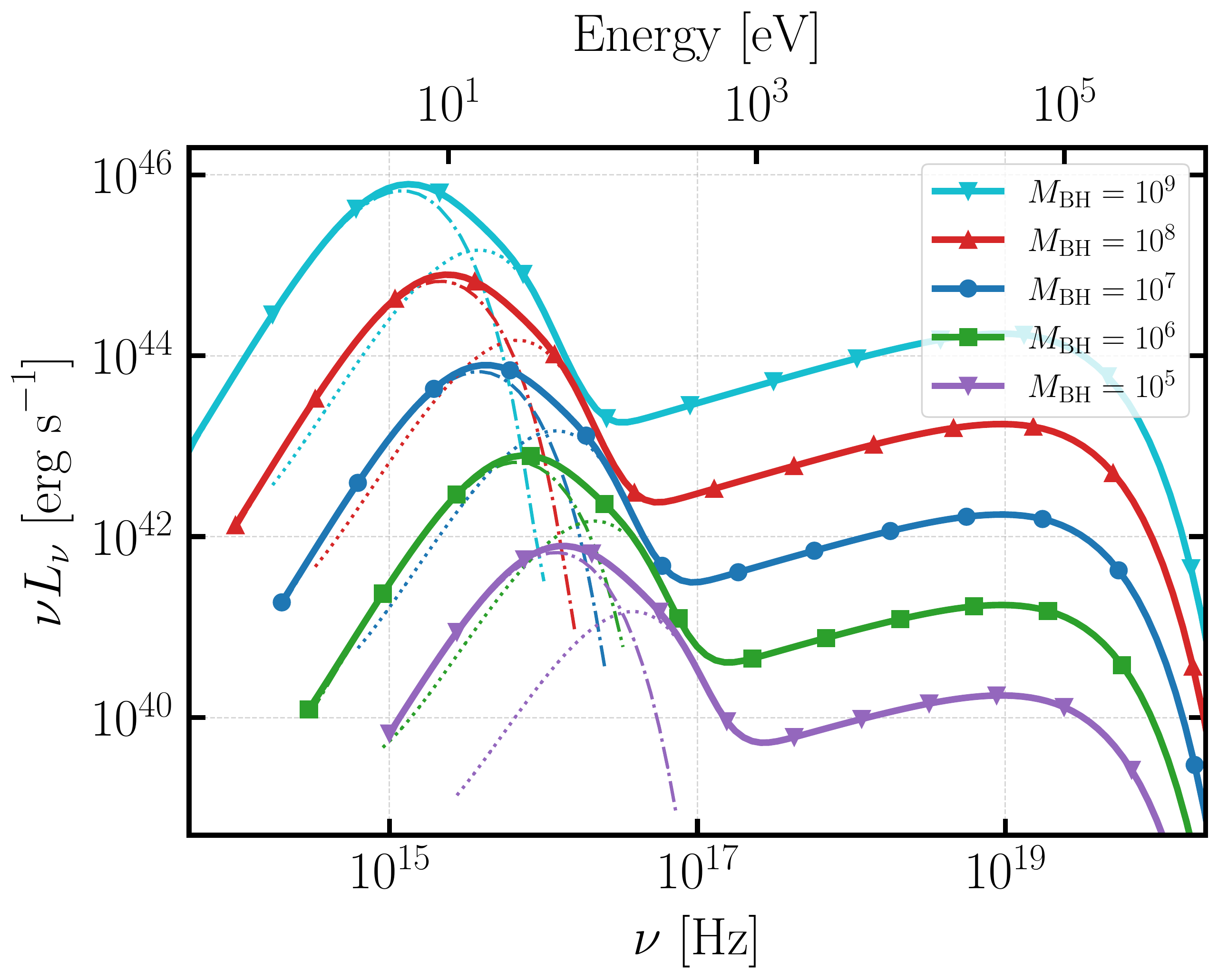}
    \caption{Spectra for various masses for \Ss simulation at a $t=3290M$.}
    \label{fig:spec_masses}
\end{figure}

\begin{figure}
    \centering
    \includegraphics[width=0.99\linewidth]{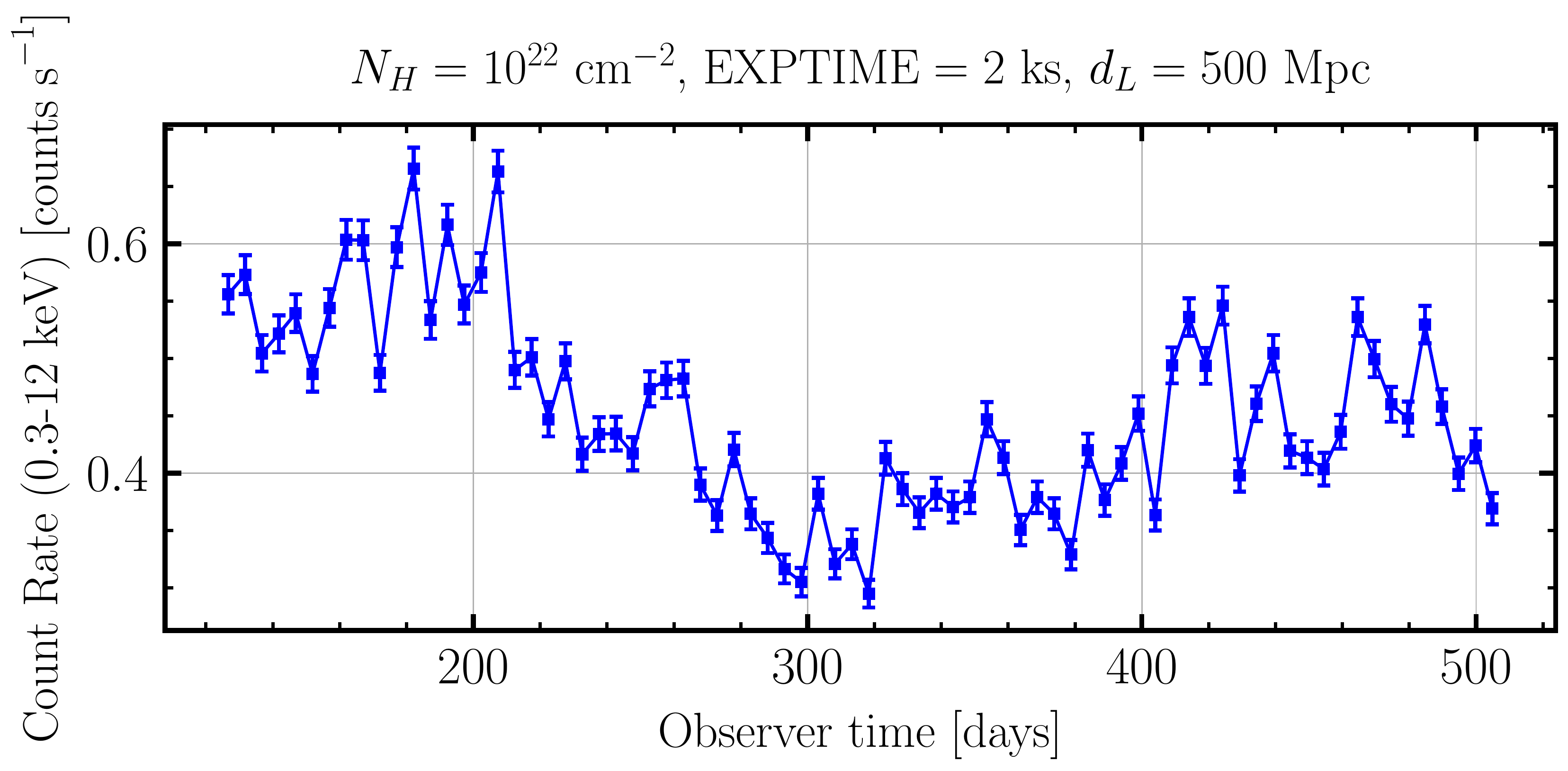}
    \caption{Simulated light curve for a sample of NICER observations of $2$ ks every $\sim 5$ days in the band ($0.3-12$ keV) for a putative SMBBH system with a mass of $10^9M_\odot$ at a distance of $500{\rm Mpc}$. We assume an absorption of $N_H=10^{22}~{\rm cm}^{-2}$. The data are taken from \Ss.}
    \label{fig:lc_X}
\end{figure}

We have explored various features of the spectrum and its time variability for a binary black hole system with a total mass of $M=10^6M_\odot$. We chose this value considering that mergers of these systems are future LISA targets and thus candidates for future multimessenger observations. However, because an AGN's luminosity is expected to scale with the black hole mass, heavier SMBBHs may be detected via their EM emission at greater distances. Additionally, as mass increases, the peak of the thermal spectrum shifts to lower energies, allowing optical/UV telescopes to observe them. As an illustration, Figure~\ref{fig:spec_masses} shows the SED from \Ss at $t=3290M$ for five different the total masses: $M=10^5$, $10^6$, $10^7$, $10^8$, and $10^9M_\odot$. The total luminosity scales as $L \propto M$, whereas the thermal peak scales as $\nu_{\rm peak} \propto M^{-0.25}$. The orbital period is also $\propto M$ and, at the separations considered, ranges from $\sim$ few minutes for $M=10^5$ to $> 30$ days for $M=10^9$. 
This large range of timescales implies the need for different observational strategies to detect SMBBHs with different masses. On the other hand, we must not only detect SMBBHs but also distinguish them from the many more single AGNs in the universe. The best way to accomplish this is to identify distinct periodic modulations in the emission.

Using the results of the \Ss simulation, we investigate two scenarios: one with a total mass of $10^9M_\odot$ and another with a mass of $10^6M_\odot$.
In the first case, the two periodic signals we identified at $\sim 0.2f_\mathcal{B}$ and $\sim 1.4f_\mathcal{B}$ would have periods of $\sim 150$ days and $ \sim 20$ days, respectively. If such a source is identified as an SMBBH candidate, a follow-up observational campaign with observations every $< 10$ days may detect these variabilities and help to confirm the nature of the source. This could be accomplished using NICER\footnote{\url{https://heasarc.gsfc.nasa.gov/docs/nicer/}}; since the NICER X-ray instrument is on board the International Space Station in a low Earth orbit, depending on the visibility, it can monitor X-ray sources up to $16$ times per day. In Figure~\ref{fig:lc_X}, we show a simulated light curve obtained by convolving our theoretical prediction with response matrices from NICER using the software {\tt XSPEC} \citep{arnaud1996}. We simulate a scenario in which recurrent observations of $2$ ks are made every $\sim 5$ days. We assume a distance of $500~{\rm Mpc}$ from the source and a typical Hydrogen column density of $N_H=10^{22} ~ {\rm cm}^{-2}$. To detect a source of this type at a farther distance, longer exposure times would be required.

On the other hand, if the total mass of the system is $10^6M_\odot$, the total duration of our simulation corresponds to $10$ hr. This period can be covered by a single observation, which should be divided appropriately to account for the light curve's variability. Athena\footnote{\url{https://sci.esa.int/web/athena}},
which will be launched in the next decade, is the best future X-ray facility for studying this type of source. Athena's effective area will be orders of magnitude larger than those of today's X-ray observatories. Assuming once more a hydrogen column density of $N_H=10^{22}~{\rm cm}^{-2}$, the variabilities could be discerned up to a distance of $d\sim 50~{\rm Mpc}$.

With regard to this timing analysis, the nonspinning situation would be very similar, with the only difference being that the signal at $\sim 1.4f_\mathcal{B}$ would be stronger than in the spinning case.

A further question concerns the prospects for observing EM emission during the merger itself. The luminosity of this event is proportional to the amount of gas that may be heated during the merger. The total mass contained within the cavity at the end of our simulations provides an upper bound on this quantity. We estimate this by taking an average of the mini-disks' mass during the final $2000M$ of each simulation. For a total black hole mass of $10^6 M_\odot$ and an accretion rate of $0.25 \dot{M}_{\rm Edd}$, the average mass in the cavity is $\sim 5.7 \times 10^{25}~{\rm g}$ and $3.6 \times 10^{25}~{\rm g}$ for \Ss and \Ns simulations, respectively. These two values are well-fitted by the following phenomenological expression:
\begin{equation}
M_{\rm cav} \lesssim 1.42 \times 10^{26} \left( \frac{M_{\rm BH}}{10^6M_\odot} \right)^2 \left( \frac{\dot{M}}{\dot{M}_{\rm Edd}} \right)\left( 1+\chi \right)~{\rm g}.
    \label{eq:mass_merger}
\end{equation}
The mass and accretion rate dependency in Eq. \ref{eq:mass_merger} are directly derived from the unit conversion between the numerical and the physical mass units, whereas the spin dependence is the simplest linear fit for the two values considered ($0$ and $0.6$) and should not be extrapolated directly to other spin values.
For an optimistic scenario involving a $10^9M_\odot$ SMBBH system accreting at a rate of $\sim \dot{M}_{\rm Edd}$, Eq. \ref{eq:mass_merger} yields a mass upper bound of $M_{\rm cav} \sim 1.5 \times 10^{32}~{\rm g}$, which corresponds to a maximum possible energy $\sim 10^{53}~{\rm erg}$, $\sim 10^4 \times$ the total radiated energy in a supernova.

\section{Conclusions}    \label{sec:conclusions}

In this work, we have analyzed the EM emission from spinning and nonspinning equal-mass SMBBHs approaching merger. We post-processed data from GRMHD simulations that evolved the system for more than $10$ orbits, and calculated images, spectra, and light curves. We have identified the primary EM signatures that may help in distinguishing a scenario involving two supermassive BHs of equal mass at separations of $\lesssim 20M$ from a scenario involving a single AGN. With the SEDs at each time, we have produced light curves at specific frequencies that reflect the behavior of the various subcomponents of the system and may be of interest for observations. Our analysis focused on face-on emission in order to identify EM signatures that are directly related to the system's dynamics and would be present regardless of the angle of view. The following are the primary outcomes of our work:

\begin{itemize}
    \item  Mini-disks just prior to the merger are, on average, $\sim 3-5$ times brighter for spinning black holes with $\chi_1=\chi_2=0.6$ than for nonspinning black holes. This translates into an increased luminosity in the far-UV and soft X-ray bands. Late in the inspiral, the contribution of the mini-disks to the thermal spectrum in \Ns is almost negligible, whereas in \Ss, it is still sustained. These differences are due to the higher radiative efficiency of the mini-disks in \Ss, which, in turn, is caused by a greater proportion of accretion occurring via circular orbits.
    \item Mini-disks in these conditions differ from NT disks onto single black holes. Normalized to the accretion rate passing through them, the mini-disks' average (face-on) radiative efficiency ranges from $\sim 1.4 \%$ for Schwarzschild black holes to $1.9 \%$ for Kerr black holes with normalized spin $\chi=0.6$. For the spinning (nonspinning) case, this value is $\sim 55 \%$ ($\sim 90 \%$) of what the NT model predicts for the same spin and disk extension. The main reason for the mini-disks' lower radiative efficiency (especially the thermal one) is that the material brought to them has so little angular momentum that it can fall into the black hole without suffering very much stress; the association of dissipation with stress implies that heating is likewise relatively meager.
The global efficiency, i.e., the ratio of total luminosity (CBD + streams + mini-disks) to CBD accretion rate, is $\sim 4 \%$ regardless of the spin of the black holes.
    \item The SED from an accretion disk onto an SMBBH of the type analyzed in this work is different from that of a single black hole disk under the same conditions. In the binary scenario, the peak of the spectrum is shifted to lower frequencies with respect to that of the single black hole disk, and the shape of the spectrum at frequencies above the peak is a broken power law rather than a single decreasing exponential. In addition, the coronal luminosity is lower due to the smaller accretion rate and radiative efficiency in the inner cavity.
    \item The emission displays periodic signals associated with the lump's dynamics. The most significant ones are at $\sim 0.2 f_\mathcal{B}$ and $\sim 1.44 f_\mathcal{B}$, which correspond to the radial oscillations of the lump and twice the beat frequency, respectively. The first signal is predicted to be present at various wavelengths, whereas the second is more pronounced in X-rays and for nonspinning black holes.
    \item Depending on the total mass of the system, periodicity in the emission during the approach to the merger may be detected in the X-ray band using various observational strategies. For high masses ($\sim 10^9M_\odot$), short observations every couple of days would be sufficient to identify the variability, assuming the source is detectable. For low masses ($\sim 10^6M_\odot$), a single observation of $>10^4$ s would detect the periodicities, but the source would need to be closer.

\end{itemize}

In upcoming work, we will investigate the variability features associated with relativistic effects by exploring low accretion rates and different viewing angles.

\software{ \texttt{numpy} \citep{numpy}, 
\texttt{XSPEC} \citep{arnaud1996},
\texttt{astropy} \citep{astropy2013}, \texttt{scipy} \citep{scipy}, and \texttt{matplotlib} \citep{matplotlib} }

\acknowledgments

We thank Jes\'us Rueda-Becerril, Mark Avara, Katie Porter, and Gustavo E. Romero for enriching discussions about SMBBH dynamics and radiative processes. E.M.G. and L.C acknowledge support from a CONICET (Argentina's National Research Council) fellowship, and from AST-2009330, AST1028087, AST-1516150, and the Frontier of Gravitational Wave Fellowship program of the RIT's Center for Computational Relativity and Gravitation.
F.L.A. and M.C. acknowledge support from NSF grants AST-2009330, AST1028087, AST-1516150, PHY-1707946, PHY-2110338, and NASA TCAN grant (NNH17ZDA001N).
S.C.N. was supported by AST-1028087, AST-1515982, and OAC-1515969, and by an appointment to the NASA Postdoctoral Program at the Goddard Space Flight Center administrated by USRA through a contract with NASA. J.H.K. was partially supported by PHY-1707826 and AST-2009260. F.G. is a CONICET researcher and acknowledges support by PIP 0113 (CONICET). This work received financial support from PICT-2017-2865 (ANPCyT). Computational resources were provided by the TACC's Frontera supercomputer allocation Nos. PHY-20010 and AST-20021. Additional resources were provided by the RIT's BlueSky and Green Pairie Clusters acquired with NSF grants AST-1028087, PHY-0722703, PHY-1229173, and PHY-1726215.




\bibliography{bhm_references}

\begin{thebibliography}{}
\expandafter\ifx\csname natexlab\endcsname\relax\def\natexlab#1{#1}\fi
\providecommand{\url}[1]{\href{#1}{#1}}
\providecommand{\dodoi}[1]{doi:~\href{http://doi.org/#1}{\nolinkurl{#1}}}
\providecommand{\doeprint}[1]{\href{http://ascl.net/#1}{\nolinkurl{http://ascl.net/#1}}}
\providecommand{\doarXiv}[1]{\href{https://arxiv.org/abs/#1}{\nolinkurl{https://arxiv.org/abs/#1}}}

\bibitem[{{Abbott} {et~al.}(2016){Abbott}, {Abbott}, {Abbott}, {Abernathy},
  {Acernese}, {Ackley}, {Adams}, {Adams}, {Addesso}, {Adhikari}, \&
  et~al.}]{GWdetect}
{Abbott}, B.~P., {Abbott}, R., {Abbott}, T.~D., {et~al.} 2016, Physical Review
  Letters, 116, 061102, \dodoi{10.1103/PhysRevLett.116.061102}

\bibitem[{Abbott {et~al.}(2017)Abbott, Abbott, Abbott, Acernese, Ackley, Adams,
  Adams, Addesso, Adhikari, Adya, {et~al.}}]{abbott2017gw170817}
Abbott, B.~P., Abbott, R., Abbott, T., {et~al.} 2017, Physical review letters,
  119, 161101

\bibitem[{Abbott {et~al.}(2021)Abbott, Abbott, Acernese, Ackley, Adams,
  Adhikari, Adhikari, Adya, Affeldt, Agarwal, {et~al.}}]{abbott2021gwtc}
Abbott, R., Abbott, T., Acernese, F., {et~al.} 2021, arXiv preprint
  arXiv:2111.03606

\bibitem[{{Alam} {et~al.}(2020){Alam}, {Arzoumanian}, {Baker}, {Blumer},
  {Bohler}, {Brazier}, {Brook}, {Burke-Spolaor}, {Caballero}, {Camuccio},
  {Chamberlain}, {Chatterjee}, {Cordes}, {Cornish}, {Crawford}, {Cromartie},
  {DeCesar}, {Demorest}, {Dolch}, {Ellis}, {Ferdman}, {Ferrara}, {Fiore},
  {Fonseca}, {Garcia}, {Garver-Daniels}, {Gentile}, {Good}, {Gusdorff},
  {Halmrast}, {Hazboun}, {Islo}, {Jennings}, {Jessup}, {Jones}, {Kaiser},
  {Kaplan}, {Kelley}, {Shapiro Key}, {Lam}, {Lazio}, {Lorimer}, {Luo}, {Lynch},
  {Madison}, {Maraccini}, {McLaughlin}, {Mingarelli}, {Ng}, {Nguyen}, {Nice},
  {Pennucci}, {Pol}, {Ramette}, {Ransom}, {Ray}, {Shapiro-Albert}, {Siemens},
  {Simon}, {Spiewak}, {Stairs}, {Stinebring}, {Stovall}, {Swiggum}, {Taylor},
  {Tripepi}, {Vallisneri}, {Vigeland}, {Witt}, \& {Zhu}}]{Nanograv2020}
{Alam}, M.~F., {Arzoumanian}, Z., {Baker}, P.~T., {et~al.} 2020, arXiv
  e-prints, arXiv:2005.06495.
\newblock \doarXiv{2005.06495}

\bibitem[{{Amaro-Seoane} {et~al.}(2017){Amaro-Seoane}, {Audley}, {Babak},
  {Baker}, {Barausse}, {Bender}, {Berti}, {Binetruy}, {Born}, {Bortoluzzi},
  {Camp}, {Caprini}, {Cardoso}, {Colpi}, {Conklin}, {Cornish}, {Cutler},
  {Danzmann}, {Dolesi}, {Ferraioli}, {Ferroni}, {Fitzsimons}, {Gair}, {Gesa
  Bote}, {Giardini}, {Gibert}, {Grimani}, {Halloin}, {Heinzel}, {Hertog},
  {Hewitson}, {Holley-Bockelmann}, {Hollington}, {Hueller}, {Inchauspe},
  {Jetzer}, {Karnesis}, {Killow}, {Klein}, {Klipstein}, {Korsakova}, {Larson},
  {Livas}, {Lloro}, {Man}, {Mance}, {Martino}, {Mateos}, {McKenzie},
  {McWilliams}, {Miller}, {Mueller}, {Nardini}, {Nelemans}, {Nofrarias},
  {Petiteau}, {Pivato}, {Plagnol}, {Porter}, {Reiche}, {Robertson},
  {Robertson}, {Rossi}, {Russano}, {Schutz}, {Sesana}, {Shoemaker}, {Slutsky},
  {Sopuerta}, {Sumner}, {Tamanini}, {Thorpe}, {Troebs}, {Vallisneri},
  {Vecchio}, {Vetrugno}, {Vitale}, {Volonteri}, {Wanner}, {Ward}, {Wass},
  {Weber}, {Ziemer}, \& {Zweifel}}]{LISA2017}
{Amaro-Seoane}, P., {Audley}, H., {Babak}, S., {et~al.} 2017, arXiv e-prints,
  arXiv:1702.00786.
\newblock \doarXiv{1702.00786}

\bibitem[{Armengol {et~al.}(2021)Armengol, Combi, Campanelli, Noble, Krolik,
  Bowen, Avara, Mewes, \& Nakano}]{lopezarmengol2021}
Armengol, F. G.~L., Combi, L., Campanelli, M., {et~al.} 2021, The Astrophysical
  Journal, 913, 16, \dodoi{10.3847/1538-4357/abf0af}

\bibitem[{{Arnaud}(1996)}]{arnaud1996}
{Arnaud}, K.~A. 1996, in Astronomical Society of the Pacific Conference Series,
  Vol. 101, Astronomical Data Analysis Software and Systems V, ed. G.~H.
  {Jacoby} \& J.~{Barnes}, 17

\bibitem[{{Astropy Collaboration} {et~al.}(2013){Astropy Collaboration},
  {Robitaille}, {Tollerud}, {Greenfield}, {Droettboom}, {Bray}, {Aldcroft},
  {Davis}, {Ginsburg}, {Price-Whelan}, {Kerzendorf}, {Conley}, {Crighton},
  {Barbary}, {Muna}, {Ferguson}, {Grollier}, {Parikh}, {Nair}, {Unther},
  {Deil}, {Woillez}, {Conseil}, {Kramer}, {Turner}, {Singer}, {Fox}, {Weaver},
  {Zabalza}, {Edwards}, {Azalee Bostroem}, {Burke}, {Casey}, {Crawford},
  {Dencheva}, {Ely}, {Jenness}, {Labrie}, {Lim}, {Pierfederici}, {Pontzen},
  {Ptak}, {Refsdal}, {Servillat}, \& {Streicher}}]{astropy2013}
{Astropy Collaboration}, {Robitaille}, T.~P., {Tollerud}, E.~J., {et~al.} 2013,
  \aap, 558, A33, \dodoi{10.1051/0004-6361/201322068}

\bibitem[{{Babak} {et~al.}(2016){Babak}, {Petiteau}, {Sesana}, {Brem},
  {Rosado}, {Taylor}, {Lassus}, {Hessels}, {Bassa}, {Burgay}, {Caballero},
  {Champion}, {Cognard}, {Desvignes}, {Gair}, {Guillemot}, {Janssen},
  {Karuppusamy}, {Kramer}, {Lazarus}, {Lee}, {Lentati}, {Liu}, {Mingarelli},
  {Os{\l}owski}, {Perrodin}, {Possenti}, {Purver}, {Sanidas}, {Smits},
  {Stappers}, {Theureau}, {Tiburzi}, {van Haasteren}, {Vecchio}, \&
  {Verbiest}}]{EPTA2016}
{Babak}, S., {Petiteau}, A., {Sesana}, A., {et~al.} 2016, \mnras, 455, 1665,
  \dodoi{10.1093/mnras/stv2092}

\bibitem[{{Begelman} {et~al.}(1980){Begelman}, {Blandford}, \&
  {Rees}}]{Begelman1980}
{Begelman}, M.~C., {Blandford}, R.~D., \& {Rees}, M.~J. 1980, \nat, 287, 307,
  \dodoi{10.1038/287307a0}

\bibitem[{{Beloborodov} \& {Illarionov}(2001)}]{BI2001}
{Beloborodov}, A.~M., \& {Illarionov}, A.~F. 2001, \mnras, 323, 167,
  \dodoi{10.1046/j.1365-8711.2001.04133.x}

\bibitem[{Blecha {et~al.}(2016)Blecha, Sijacki, Kelley, Torrey, Vogelsberger,
  Nelson, Springel, Snyder, \& Hernquist}]{blecha2016recoiling}
Blecha, L., Sijacki, D., Kelley, L.~Z., {et~al.} 2016, Monthly Notices of the
  Royal Astronomical Society, 456, 961

\bibitem[{Bogdanovic {et~al.}(2021)Bogdanovic, Miller, \&
  Blecha}]{bogdanovic2021electromagnetic}
Bogdanovic, T., Miller, M.~C., \& Blecha, L. 2021, arXiv preprint
  arXiv:2109.03262

\bibitem[{{Bowen} {et~al.}(2017){Bowen}, {Campanelli}, {Krolik}, {Mewes}, \&
  {Noble}}]{Bowen17}
{Bowen}, D.~B., {Campanelli}, M., {Krolik}, J.~H., {Mewes}, V., \& {Noble},
  S.~C. 2017, \apj, 838, 42, \dodoi{10.3847/1538-4357/aa63f3}

\bibitem[{Bowen {et~al.}(2018)Bowen, Mewes, Campanelli, Noble, Krolik, \&
  Zilh{\~a}o}]{Bowen2018}
Bowen, D.~B., Mewes, V., Campanelli, M., {et~al.} 2018, The Astrophysical
  Journal Letters, 853, L17

\bibitem[{{Bowen} {et~al.}(2019){Bowen}, {Mewes}, {Noble}, {Avara},
  {Campanelli}, \& {Krolik}}]{Bowen2019}
{Bowen}, D.~B., {Mewes}, V., {Noble}, S.~C., {et~al.} 2019, \apj, 879, 76,
  \dodoi{10.3847/1538-4357/ab2453}

\bibitem[{{Bronzwaer} \& {Falcke}(2021)}]{bronzwaer2021}
{Bronzwaer}, T., \& {Falcke}, H. 2021, \apj, 920, 155,
  \dodoi{10.3847/1538-4357/ac1738}

\bibitem[{Campanelli {et~al.}(2006)Campanelli, Lousto, \&
  Zlochower}]{Campanelli:2006uy}
Campanelli, M., Lousto, C.~O., \& Zlochower, Y. 2006, Phys. Rev., D74, 041501,
  \dodoi{10.1103/PhysRevD.74.041501}

\bibitem[{Campanelli {et~al.}(2007)Campanelli, Lousto, Zlochower, \&
  Merritt}]{Campanelli:2007apj}
Campanelli, M., Lousto, C.~O., Zlochower, Y., \& Merritt, D. 2007, Astrophys.
  J., 659, L5, \dodoi{10.1086/516712}

\bibitem[{Cattorini {et~al.}(2021)Cattorini, Giacomazzo, Haardt, \&
  Colpi}]{cattorini2021}
Cattorini, F., Giacomazzo, B., Haardt, F., \& Colpi, M. 2021, Physical Review
  D, 103, 103022

\bibitem[{Chapon {et~al.}(2013)Chapon, Mayer, \& Teyssier}]{Chapon2013}
Chapon, D., Mayer, L., \& Teyssier, R. 2013, Monthly Notices of the Royal
  Astronomical Society, 429, 3114, \dodoi{10.1093/mnras/sts568}

\bibitem[{Charisi {et~al.}(2018)Charisi, Haiman, Schiminovich, \&
  D'Orazio}]{charisi2018testing}
Charisi, M., Haiman, Z., Schiminovich, D., \& D'Orazio, D.~J. 2018, Monthly
  Notices of the Royal Astronomical Society, 476, 4617

\bibitem[{{Combi} {et~al.}(2021{\natexlab{a}}){Combi}, {Armengol},
  {Campanelli}, {Ireland}, {Noble}, {Nakano}, \& {Bowen}}]{Combi2021a}
{Combi}, L., {Armengol}, F. G.~L., {Campanelli}, M., {et~al.}
  2021{\natexlab{a}}, \prd, 104, 044041, \dodoi{10.1103/PhysRevD.104.044041}

\bibitem[{{Combi} {et~al.}(2021{\natexlab{b}}){Combi}, {Lopez Armengol},
  {Campanelli}, {Noble}, {Avara}, {Krolik}, \& {Bowen}}]{Combi2021b}
{Combi}, L., {Lopez Armengol}, F.~G., {Campanelli}, M., {et~al.}
  2021{\natexlab{b}}, arXiv e-prints, arXiv:2109.01307.
\newblock \doarXiv{2109.01307}

\bibitem[{{d'Ascoli} {et~al.}(2018){d'Ascoli}, {Noble}, {Bowen}, {Campanelli},
  {Krolik}, \& {Mewes}}]{dAscoli2018}
{d'Ascoli}, S., {Noble}, S.~C., {Bowen}, D.~B., {et~al.} 2018, \apj, 865, 140,
  \dodoi{10.3847/1538-4357/aad8b4}

\bibitem[{{Davelaar} \& {Haiman}(2021{\natexlab{a}})}]{davelaar2021a}
{Davelaar}, J., \& {Haiman}, Z. 2021{\natexlab{a}}, arXiv e-prints,
  arXiv:2112.05828.
\newblock \doarXiv{2112.05828}

\bibitem[{{Davelaar} \& {Haiman}(2021{\natexlab{b}})}]{davelaar2021b}
---. 2021{\natexlab{b}}, arXiv e-prints, arXiv:2112.05829.
\newblock \doarXiv{2112.05829}

\bibitem[{{Davis} {et~al.}(2005){Davis}, {Blaes}, {Hubeny}, \&
  {Turner}}]{SDavis2005}
{Davis}, S.~W., {Blaes}, O.~M., {Hubeny}, I., \& {Turner}, N.~J. 2005, \apj,
  621, 372, \dodoi{10.1086/427278}

\bibitem[{{Derdzinski} {et~al.}(2021){Derdzinski}, {D'Orazio}, {Duffell},
  {Haiman}, \& {MacFadyen}}]{Derdzinski2021}
{Derdzinski}, A., {D'Orazio}, D., {Duffell}, P., {Haiman}, Z., \& {MacFadyen},
  A. 2021, \mnras, 501, 3540, \dodoi{10.1093/mnras/staa3976}

\bibitem[{{Derdzinski} {et~al.}(2019){Derdzinski}, {D'Orazio}, {Duffell},
  {Haiman}, \& {MacFadyen}}]{Derdzinski2019}
{Derdzinski}, A.~M., {D'Orazio}, D., {Duffell}, P., {Haiman}, Z., \&
  {MacFadyen}, A. 2019, \mnras, 486, 2754, \dodoi{10.1093/mnras/stz1026}

\bibitem[{{D'Orazio} \& {Di Stefano}(2018)}]{Dorazio2018}
{D'Orazio}, D.~J., \& {Di Stefano}, R. 2018, \mnras, 474, 2975,
  \dodoi{10.1093/mnras/stx2936}

\bibitem[{{D'Orazio} {et~al.}(2016){D'Orazio}, {Haiman}, {Duffell},
  {MacFadyen}, \& {Farris}}]{DOrazio2016}
{D'Orazio}, D.~J., {Haiman}, Z., {Duffell}, P., {MacFadyen}, A., \& {Farris},
  B. 2016, \mnras, 459, 2379, \dodoi{10.1093/mnras/stw792}

\bibitem[{{D'Orazio} {et~al.}(2013){D'Orazio}, {Haiman}, \&
  {MacFadyen}}]{DOrazio13}
{D'Orazio}, D.~J., {Haiman}, Z., \& {MacFadyen}, A. 2013, Monthly Notices of
  the Royal Astronomical Society, 436, 2997, \dodoi{10.1093/mnras/stt1787}

\bibitem[{{D'Orazio} {et~al.}(2015){D'Orazio}, {Haiman}, \&
  {Schiminovich}}]{dorazio2015nature}
{D'Orazio}, D.~J., {Haiman}, Z., \& {Schiminovich}, D. 2015, \nat, 525, 351,
  \dodoi{10.1038/nature15262}

\bibitem[{{Dotti} {et~al.}(2007){Dotti}, {Colpi}, {Haardt}, \&
  {Mayer}}]{Dotti2007}
{Dotti}, M., {Colpi}, M., {Haardt}, F., \& {Mayer}, L. 2007, \mnras, 379, 956,
  \dodoi{10.1111/j.1365-2966.2007.12010.x}

\bibitem[{Dotti {et~al.}(2009)Dotti, Ruszkowski, Paredi, Colpi, Volonteri, \&
  Haardt}]{Dotti2009b}
Dotti, M., Ruszkowski, M., Paredi, L., {et~al.} 2009, Monthly Notices of the
  Royal Astronomical Society, 396, 1640,
  \dodoi{10.1111/j.1365-2966.2009.14840.x}

\bibitem[{{Duffell} {et~al.}(2020){Duffell}, {D'Orazio}, {Derdzinski},
  {Haiman}, {MacFadyen}, {Rosen}, \& {Zrake}}]{Duffell2020}
{Duffell}, P.~C., {D'Orazio}, D., {Derdzinski}, A., {et~al.} 2020, \apj, 901,
  25, \dodoi{10.3847/1538-4357/abab95}

\bibitem[{{Escala} {et~al.}(2004){Escala}, {Larson}, {Coppi}, \&
  {Mardones}}]{Escala2004}
{Escala}, A., {Larson}, R.~B., {Coppi}, P.~S., \& {Mardones}, D. 2004, \apj,
  607, 765, \dodoi{10.1086/386278}

\bibitem[{{Escala} {et~al.}(2005){Escala}, {Larson}, {Coppi}, \&
  {Mardones}}]{Escala2005}
---. 2005, \apj, 630, 152, \dodoi{10.1086/431747}

\bibitem[{Farris {et~al.}(2014{\natexlab{a}})Farris, Duffell, MacFadyen, \&
  Haiman}]{Farris2014a}
Farris, B.~D., Duffell, P., MacFadyen, A.~I., \& Haiman, Z. 2014{\natexlab{a}},
  The Astrophysical Journal, 783, 134, \dodoi{10.1088/0004-637x/783/2/134}

\bibitem[{Farris {et~al.}(2014{\natexlab{b}})Farris, Duffell, MacFadyen, \&
  Haiman}]{Farris2014b}
---. 2014{\natexlab{b}}, Monthly Notices of the Royal Astronomical Society:
  Letters, 446, L36, \dodoi{10.1093/mnrasl/slu160}

\bibitem[{{Farris} {et~al.}(2014){Farris}, {Duffell}, {MacFadyen}, \&
  {Haiman}}]{Farris14}
{Farris}, B.~D., {Duffell}, P., {MacFadyen}, A.~I., \& {Haiman}, Z. 2014, The
  Astrophysical Journal, 783, 134, \dodoi{10.1088/0004-637X/783/2/134}

\bibitem[{{Farris} {et~al.}(2015{\natexlab{a}}){Farris}, {Duffell},
  {MacFadyen}, \& {Haiman}}]{Farris15a}
---. 2015{\natexlab{a}}, \mnras, 446, L36, \dodoi{10.1093/mnrasl/slu160}

\bibitem[{{Farris} {et~al.}(2015{\natexlab{b}}){Farris}, {Duffell},
  {MacFadyen}, \& {Haiman}}]{Farris15b}
---. 2015{\natexlab{b}}, \mnras, 447, L80, \dodoi{10.1093/mnrasl/slu184}

\bibitem[{{Farris} {et~al.}(2011){Farris}, {Liu}, \& {Shapiro}}]{Farris11}
{Farris}, B.~D., {Liu}, Y.~T., \& {Shapiro}, S.~L. 2011, \prd, 84, 024024,
  \dodoi{10.1103/PhysRevD.84.024024}

\bibitem[{Giacomazzo {et~al.}(2012)Giacomazzo, Baker, Miller, Reynolds, \& van
  Meter}]{giacomazzo2012}
Giacomazzo, B., Baker, J.~G., Miller, M.~C., Reynolds, C.~S., \& van Meter,
  J.~R. 2012, The Astrophysical Journal Letters, 752, L15

\bibitem[{Gold {et~al.}(2014)Gold, Paschalidis, Etienne, Shapiro, \&
  Pfeiffer}]{Gold14}
Gold, R., Paschalidis, V., Etienne, Z.~B., Shapiro, S.~L., \& Pfeiffer, H.~P.
  2014, Phys. Rev. D, 89, 064060, \dodoi{10.1103/PhysRevD.89.064060}

\bibitem[{{Gold} {et~al.}(2014){Gold}, {Paschalidis}, {Ruiz}, {Shapiro},
  {Etienne}, \& {Pfeiffer}}]{Gold2014b}
{Gold}, R., {Paschalidis}, V., {Ruiz}, M., {et~al.} 2014, \prd, 90, 104030,
  \dodoi{10.1103/PhysRevD.90.104030}

\bibitem[{Graham {et~al.}(2015)Graham, Djorgovski, Stern, Drake, Mahabal,
  Donalek, Glikman, Larson, \& Christensen}]{graham2015systematic}
Graham, M.~J., Djorgovski, S., Stern, D., {et~al.} 2015, Monthly Notices of the
  Royal Astronomical Society, 453, 1562

\bibitem[{{Gralla}(2021)}]{Gralla2021}
{Gralla}, S.~E. 2021, \prd, 103, 024023, \dodoi{10.1103/PhysRevD.103.024023}

\bibitem[{Harris {et~al.}(2020)Harris, Millman, van~der Walt, Gommers,
  Virtanen, Cournapeau, Wieser, Taylor, Berg, Smith, Kern, Picus, Hoyer, van
  Kerkwijk, Brett, Haldane, del R{'{\i}}o, Wiebe, Peterson,
  G{'{e}}rard-Marchant, Sheppard, Reddy, Weckesser, Abbasi, Gohlke, \&
  Oliphant}]{numpy}
Harris, C.~R., Millman, K.~J., van~der Walt, S.~J., {et~al.} 2020, Nature, 585,
  357, \dodoi{10.1038/s41586-020-2649-2}

\bibitem[{Hu {et~al.}(2020)Hu, D’Orazio, Haiman, Smith, Snios, Charisi, \&
  Di~Stefano}]{hu2020spikey}
Hu, B.~X., D’Orazio, D.~J., Haiman, Z., {et~al.} 2020, Monthly Notices of the
  Royal Astronomical Society, 495, 4061

\bibitem[{Hunter(2007)}]{matplotlib}
Hunter, J.~D. 2007, Computing in Science \& Engineering, 9, 90,
  \dodoi{10.1109/MCSE.2007.55}

\bibitem[{Ingram {et~al.}(2021)Ingram, Motta, Aigrain, \&
  Karastergiou}]{ingram2021self}
Ingram, A., Motta, S.~E., Aigrain, S., \& Karastergiou, A. 2021, Monthly
  Notices of the Royal Astronomical Society, 503, 1703

\bibitem[{Kelley {et~al.}(2019)Kelley, Haiman, Sesana, \&
  Hernquist}]{kelley2019massive}
Kelley, L.~Z., Haiman, Z., Sesana, A., \& Hernquist, L. 2019, Monthly Notices
  of the Royal Astronomical Society, 485, 1579

\bibitem[{Kelly {et~al.}(2017)Kelly, Baker, Etienne, Giacomazzo, \&
  Schnittman}]{Kelly2017}
Kelly, B.~J., Baker, J.~G., Etienne, Z.~B., Giacomazzo, B., \& Schnittman, J.
  2017, Phys. Rev. D, 96, 123003, \dodoi{10.1103/PhysRevD.96.123003}

\bibitem[{{Kelly} {et~al.}(2021){Kelly}, {Etienne}, {Golomb}, {Schnittman},
  {Baker}, {Noble}, \& {Ryan}}]{kelly2021electromagnetic}
{Kelly}, B.~J., {Etienne}, Z.~B., {Golomb}, J., {et~al.} 2021, \prd, 103,
  063039, \dodoi{10.1103/PhysRevD.103.063039}

\bibitem[{Khan {et~al.}(2019)Khan, Mirza, \& Holley-Bockelmann}]{Khan2019}
Khan, F.~M., Mirza, M.~A., \& Holley-Bockelmann, K. 2019, Monthly Notices of
  the Royal Astronomical Society, 492, 256, \dodoi{10.1093/mnras/stz3360}

\bibitem[{{Kinch} {et~al.}(2020){Kinch}, {Noble}, {Schnittman}, \&
  {Krolik}}]{Kinch2020}
{Kinch}, B.~E., {Noble}, S.~C., {Schnittman}, J.~D., \& {Krolik}, J.~H. 2020,
  \apj, 904, 117, \dodoi{10.3847/1538-4357/abc176}

\bibitem[{{Kinch} {et~al.}(2021){Kinch}, {Schnittman}, {Noble}, {Kallman}, \&
  {Krolik}}]{Kinch2021}
{Kinch}, B.~E., {Schnittman}, J.~D., {Noble}, S.~C., {Kallman}, T.~R., \&
  {Krolik}, J.~H. 2021, \apj, 922, 270, \dodoi{10.3847/1538-4357/ac2b9a}

\bibitem[{Kocsis {et~al.}(2011)Kocsis, Yunes, \& Loeb}]{Kocsis:2011dr}
Kocsis, B., Yunes, N., \& Loeb, A. 2011, Phys. Rev., D84, 024032,
  \dodoi{10.1103/PhysRevD.84.024032}

\bibitem[{{Krolik}(2010)}]{Krolik10}
{Krolik}, J.~H. 2010, \apj, 709, 774, \dodoi{10.1088/0004-637X/709/2/774}

\bibitem[{Krolik {et~al.}(2019)Krolik, Volonteri, Dubois, \&
  Devriendt}]{Krolik2019}
Krolik, J.~H., Volonteri, M., Dubois, Y., \& Devriendt, J. 2019, The
  Astrophysical Journal, 879, 110, \dodoi{10.3847/1538-4357/ab24c9}

\bibitem[{Liu {et~al.}(2003)Liu, Wu, \& Cao}]{Liu2003}
Liu, F.~K., Wu, X.-B., \& Cao, S.~L. 2003, Monthly Notices of the Royal
  Astronomical Society, 340, 411, \dodoi{10.1046/j.1365-8711.2003.06235.x}

\bibitem[{{MacFadyen} \& {Milosavljevi{\'c}}(2008)}]{MM08}
{MacFadyen}, A.~I., \& {Milosavljevi{\'c}}, M. 2008, \apj, 672, 83,
  \dodoi{10.1086/523869}

\bibitem[{Mayer {et~al.}(2007)Mayer, Kazantzidis, Madau, Colpi, Quinn, \&
  Wadsley}]{Mayer2007}
Mayer, L., Kazantzidis, S., Madau, P., {et~al.} 2007, Science, 316, 1874,
  \dodoi{10.1126/science.1141858}

\bibitem[{{Merritt}(2004)}]{Merrit2004}
{Merritt}, D. 2004, in Coevolution of Black Holes and Galaxies, ed. L.~C. {Ho},
  263.
\newblock \doarXiv{astro-ph/0301257}

\bibitem[{Merritt(2006)}]{Merritt2006}
Merritt, D. 2006, Reports on Progress in Physics, 69, 2513,
  \dodoi{10.1088/0034-4885/69/9/r01}

\bibitem[{{Milosavljevi{\'c}} \& {Phinney}(2005)}]{Milosavljevic2005}
{Milosavljevi{\'c}}, M., \& {Phinney}, E.~S. 2005, \apjl, 622, L93,
  \dodoi{10.1086/429618}

\bibitem[{{Miranda} {et~al.}(2017){Miranda}, {Mu{\~n}oz}, \&
  {Lai}}]{Miranda2017}
{Miranda}, R., {Mu{\~n}oz}, D.~J., \& {Lai}, D. 2017, \mnras, 466, 1170,
  \dodoi{10.1093/mnras/stw3189}

\bibitem[{Mirza {et~al.}(2017)Mirza, Tahir, Khan, Holley-Bockelmann, Baig,
  Berczik, \& Chishtie}]{Mirza2017}
Mirza, M.~A., Tahir, A., Khan, F.~M., {et~al.} 2017, Monthly Notices of the
  Royal Astronomical Society, 470, 940, \dodoi{10.1093/mnras/stx1248}

\bibitem[{Moesta {et~al.}(2012)Moesta, Alic, Rezzolla, Zanotti, \&
  Palenzuela}]{moesta2012detectability}
Moesta, P., Alic, D., Rezzolla, L., Zanotti, O., \& Palenzuela, C. 2012, The
  Astrophysical Journal Letters, 749, L32

\bibitem[{{Moody} {et~al.}(2019){Moody}, {Shi}, \& {Stone}}]{Moody2019}
{Moody}, M. S.~L., {Shi}, J.-M., \& {Stone}, J.~M. 2019, \apj, 875, 66,
  \dodoi{10.3847/1538-4357/ab09ee}

\bibitem[{{M{\"o}sta} {et~al.}(2019){M{\"o}sta}, {Taam}, \&
  {Duffell}}]{Mosta2019}
{M{\"o}sta}, P., {Taam}, R.~E., \& {Duffell}, P.~C. 2019, \apjl, 875, L21,
  \dodoi{10.3847/2041-8213/ab1592}

\bibitem[{{Mu{\~n}oz} \& {Lai}(2016)}]{Munoz2016}
{Mu{\~n}oz}, D.~J., \& {Lai}, D. 2016, \apj, 827, 43,
  \dodoi{10.3847/0004-637X/827/1/43}

\bibitem[{{Mu{\~n}oz} {et~al.}(2020){Mu{\~n}oz}, {Lai}, {Kratter}, \&
  {Miranda}}]{Munoz2020a}
{Mu{\~n}oz}, D.~J., {Lai}, D., {Kratter}, K., \& {Miranda}, R. 2020, \apj, 889,
  114, \dodoi{10.3847/1538-4357/ab5d33}

\bibitem[{{Mu{\~n}oz} \& {Lithwick}(2020)}]{Munoz2020b}
{Mu{\~n}oz}, D.~J., \& {Lithwick}, Y. 2020, \apj, 905, 106,
  \dodoi{10.3847/1538-4357/abc74c}

\bibitem[{{Mu{\~n}oz} {et~al.}(2019){Mu{\~n}oz}, {Miranda}, \&
  {Lai}}]{Munoz2019}
{Mu{\~n}oz}, D.~J., {Miranda}, R., \& {Lai}, D. 2019, \apj, 871, 84,
  \dodoi{10.3847/1538-4357/aaf867}

\bibitem[{Mundim {et~al.}(2014)Mundim, Nakano, Yunes, Campanelli, Noble, \&
  Zlochower}]{Mundim2014}
Mundim, B.~C., Nakano, H., Yunes, N., {et~al.} 2014, Phys. Rev. D, 89, 084008,
  \dodoi{10.1103/PhysRevD.89.084008}

\bibitem[{Noble {et~al.}(2021)Noble, Krolik, Campanelli, Zlochower, Mundim,
  Nakano, \& Zilh{\~a}o}]{noble2021}
Noble, S.~C., Krolik, J.~H., Campanelli, M., {et~al.} 2021, arXiv preprint
  arXiv:2103.12100

\bibitem[{{Noble} {et~al.}(2009){Noble}, {Krolik}, \& {Hawley}}]{Noble09}
{Noble}, S.~C., {Krolik}, J.~H., \& {Hawley}, J.~F. 2009, \apj, 692, 411,
  \dodoi{10.1088/0004-637X/692/1/411}

\bibitem[{{Noble} {et~al.}(2007){Noble}, {Leung}, {Gammie}, \&
  {Book}}]{Noble07}
{Noble}, S.~C., {Leung}, P.~K., {Gammie}, C.~F., \& {Book}, L.~G. 2007,
  Classical and Quantum Gravity, 24, 259, \dodoi{10.1088/0264-9381/24/12/S17}

\bibitem[{Noble {et~al.}(2012)Noble, Mundim, Nakano, Krolik, Campanelli,
  Zlochower, \& Yunes}]{Noble12}
Noble, S.~C., Mundim, B.~C., Nakano, H., {et~al.} 2012, The Astrophysical
  Journal, 755, 51.
\newblock \url{http://stacks.iop.org/0004-637X/755/i=1/a=51}

\bibitem[{{Novikov} \& {Thorne}(1973)}]{NovikovThorne73}
{Novikov}, I.~D., \& {Thorne}, K.~S. 1973, in Black Holes (Les Astres Occlus),
  ed. C.~{Dewitt} \& B.~S. {Dewitt}, 343--450

\bibitem[{{O'Neill} {et~al.}(2022){O'Neill}, {Kiehlmann}, {Readhead}, {Aller},
  {Blandford}, {Liodakis}, {Lister}, {Mr{\'o}z}, {O'Dea}, {Pearson}, {Ravi},
  {Vallisneri}, {Cleary}, {Graham}, {Grainge}, {Hodges}, {Hovatta},
  {L{\"a}hteenm{\"a}ki}, {Lamb}, {Lazio}, {Max-Moerbeck}, {Pavlidou}, {Prince},
  {Reeves}, {Tornikoski}, {Vergara de la Parra}, \& {Zensus}}]{ONeill2022}
{O'Neill}, S., {Kiehlmann}, S., {Readhead}, A.~C.~S., {et~al.} 2022, \apjl,
  926, L35, \dodoi{10.3847/2041-8213/ac504b}

\bibitem[{{O'Neill} {et~al.}(2009){O'Neill}, {Miller}, {Bogdanovi{\'c}},
  {Reynolds}, \& {Schnittman}}]{ONeill09}
{O'Neill}, S.~M., {Miller}, M.~C., {Bogdanovi{\'c}}, T., {Reynolds}, C.~S., \&
  {Schnittman}, J.~D. 2009, \apj, 700, 859, \dodoi{10.1088/0004-637X/700/1/859}

\bibitem[{Palenzuela {et~al.}(2010)Palenzuela, Lehner, \&
  Liebling}]{palenzuela2010dual}
Palenzuela, C., Lehner, L., \& Liebling, S.~L. 2010, Science, 329, 927

\bibitem[{{Paschalidis} {et~al.}(2021){Paschalidis}, {Bright}, {Ruiz}, \&
  {Gold}}]{paschalidis2021minidisk}
{Paschalidis}, V., {Bright}, J., {Ruiz}, M., \& {Gold}, R. 2021, \apjl, 910,
  L26, \dodoi{10.3847/2041-8213/abee21}

\bibitem[{Perna {et~al.}(2018)Perna, Chruslinska, Corsi, \&
  Belczynski}]{perna2018binary}
Perna, R., Chruslinska, M., Corsi, A., \& Belczynski, K. 2018, Monthly Notices
  of the Royal Astronomical Society, 477, 4228

\bibitem[{Pringle(1991)}]{Pringle1991x}
Pringle, J.~E. 1991, Monthly Notices of the Royal Astronomical Society, 248,
  754, \dodoi{10.1093/mnras/248.4.754}

\bibitem[{Reardon {et~al.}(2015)Reardon, Hobbs, Coles, Levin, Keith, Bailes,
  Bhat, Burke-Spolaor, Dai, Kerr, Lasky, Manchester, Os?owski, Ravi, Shannon,
  van Straten, Toomey, Wang, Wen, You, \& Zhu}]{Reardon2015x}
Reardon, D.~J., Hobbs, G., Coles, W., {et~al.} 2015, Monthly Notices of the
  Royal Astronomical Society, 455, 1751, \dodoi{10.1093/mnras/stv2395}

\bibitem[{Roedig {et~al.}(2014)Roedig, Krolik, \& Miller}]{Roedig2014}
Roedig, C., Krolik, J.~H., \& Miller, M.~C. 2014, The Astrophysical Journal,
  785, 115, \dodoi{10.1088/0004-637x/785/2/115}

\bibitem[{{Rossi} {et~al.}(2009){Rossi}, {Lodato}, {Armitage}, {Pringle}, \&
  {King}}]{Rossi09}
{Rossi}, E.~M., {Lodato}, G., {Armitage}, P.~J., {Pringle}, J.~E., \& {King},
  A.~R. 2009, ArXiv e-prints.
\newblock \doarXiv{0910.0002}

\bibitem[{{Ryan} \& {MacFadyen}(2017)}]{RyanMacFadyen17}
{Ryan}, G., \& {MacFadyen}, A. 2017, \apj, 835, 199,
  \dodoi{10.3847/1538-4357/835/2/199}

\bibitem[{{Schnittman} \& {Krolik}(2008)}]{SK08}
{Schnittman}, J.~D., \& {Krolik}, J.~H. 2008, \apj, 684, 835,
  \dodoi{10.1086/590363}

\bibitem[{{Schnittman} {et~al.}(2016){Schnittman}, {Krolik}, \&
  {Noble}}]{Schnittman2016}
{Schnittman}, J.~D., {Krolik}, J.~H., \& {Noble}, S.~C. 2016, \apj, 819, 48,
  \dodoi{10.3847/0004-637X/819/1/48}

\bibitem[{Schoenmakers {et~al.}(2000)Schoenmakers, de~Bruyn, Rottgering,
  van~der Laan, \& Kaiser}]{Shoenmakers2000}
Schoenmakers, A.~P., de~Bruyn, A.~G., Rottgering, H. J.~A., van~der Laan, H.,
  \& Kaiser, C.~R. 2000, Monthly Notices of the Royal Astronomical Society,
  315, 371, \dodoi{10.1046/j.1365-8711.2000.03430.x}

\bibitem[{{Sesana} \& {Khan}(2015)}]{SesanaKhan2015}
{Sesana}, A., \& {Khan}, F.~M. 2015, \mnras, 454, L66,
  \dodoi{10.1093/mnrasl/slv131}

\bibitem[{{Sesana} {et~al.}(2012){Sesana}, {Roedig}, {Reynolds}, \&
  {Dotti}}]{Sesana2012}
{Sesana}, A., {Roedig}, C., {Reynolds}, M.~T., \& {Dotti}, M. 2012, \mnras,
  420, 860, \dodoi{10.1111/j.1365-2966.2011.20097.x}

\bibitem[{{Shi} \& {Krolik}(2015)}]{Shi2015}
{Shi}, J.-M., \& {Krolik}, J.~H. 2015, \apj, 807, 131,
  \dodoi{10.1088/0004-637X/807/2/131}

\bibitem[{Shi {et~al.}(2012)Shi, Krolik, Lubow, \& Hawley}]{Shi2012}
Shi, J.-M., Krolik, J.~H., Lubow, S.~H., \& Hawley, J.~F. 2012, The
  Astrophysical Journal, 749, 118, \dodoi{10.1088/0004-637x/749/2/118}

\bibitem[{{Shi} {et~al.}(2012){Shi}, {Krolik}, {Lubow}, \& {Hawley}}]{Shi12}
{Shi}, J.-M., {Krolik}, J.~H., {Lubow}, S.~H., \& {Hawley}, J.~F. 2012, \apj,
  749, 118, \dodoi{10.1088/0004-637X/749/2/118}

\bibitem[{Springel {et~al.}(2005)Springel, Matteo, \& Hernquist}]{Springel2005}
Springel, V., Matteo, T.~D., \& Hernquist, L. 2005, The Astrophysical Journal,
  620, L79, \dodoi{10.1086/428772}

\bibitem[{{Tang} {et~al.}(2018){Tang}, {Haiman}, \& {MacFadyen}}]{Tang2018}
{Tang}, Y., {Haiman}, Z., \& {MacFadyen}, A. 2018, \mnras, 476, 2249,
  \dodoi{10.1093/mnras/sty423}

\bibitem[{{Tiede} {et~al.}(2020{\natexlab{a}}){Tiede}, {Zrake}, {MacFadyen}, \&
  {Haiman}}]{Tiede+2020}
{Tiede}, C., {Zrake}, J., {MacFadyen}, A., \& {Haiman}, Z. 2020{\natexlab{a}},
  \apj, 900, 43, \dodoi{10.3847/1538-4357/aba432}

\bibitem[{{Tiede} {et~al.}(2020{\natexlab{b}}){Tiede}, {Zrake}, {MacFadyen}, \&
  {Haiman}}]{Tiede2020}
---. 2020{\natexlab{b}}, \apj, 900, 43, \dodoi{10.3847/1538-4357/aba432}

\bibitem[{{Tiede} {et~al.}(2021){Tiede}, {Zrake}, {MacFadyen}, \&
  {Haiman}}]{Tiede2021}
---. 2021, arXiv e-prints, arXiv:2111.04721.
\newblock \doarXiv{2111.04721}

\bibitem[{Valtonen {et~al.}(2008)Valtonen, Lehto, Nilsson, Heidt, Takalo,
  Sillanp{\"a}{\"a}, Villforth, Kidger, Poyner, Pursimo,
  {et~al.}}]{valtonen2008massive}
Valtonen, M.~J., Lehto, H., Nilsson, K., {et~al.} 2008, Nature, 452, 851

\bibitem[{Virtanen {et~al.}(2020)Virtanen, Gommers, Oliphant, Haberland, Reddy,
  Cournapeau, Burovski, Peterson, Weckesser, Bright, {van der Walt}, Brett,
  Wilson, Millman, Mayorov, Nelson, Jones, Kern, Larson, Carey, Polat, Feng,
  Moore, {VanderPlas}, Laxalde, Perktold, Cimrman, Henriksen, Quintero, Harris,
  Archibald, Ribeiro, Pedregosa, {van Mulbregt}, \& {SciPy 1.0
  Contributors}}]{scipy}
Virtanen, P., Gommers, R., Oliphant, T.~E., {et~al.} 2020, Nature Methods, 17,
  261, \dodoi{10.1038/s41592-019-0686-2}

\bibitem[{Volonteri(2010)}]{volonteri2010formation}
Volonteri, M. 2010, The Astronomy and Astrophysics Review, 18, 279

\bibitem[{Volonteri \& Madau(2008)}]{volonteri2008off}
Volonteri, M., \& Madau, P. 2008, The Astrophysical Journal Letters, 687, L57

\bibitem[{{Westernacher-Schneider} {et~al.}(2021){Westernacher-Schneider},
  {Zrake}, {MacFadyen}, \& {Haiman}}]{westernacher2021}
{Westernacher-Schneider}, J.~R., {Zrake}, J., {MacFadyen}, A., \& {Haiman}, Z.
  2021, arXiv e-prints, arXiv:2111.06882.
\newblock \doarXiv{2111.06882}

\bibitem[{{Zhu} {et~al.}(2012){Zhu}, {Davis}, {Narayan}, {Kulkarni}, {Penna},
  \& {McClintock}}]{ZhuDavis2012}
{Zhu}, Y., {Davis}, S.~W., {Narayan}, R., {et~al.} 2012, \mnras, 424, 2504,
  \dodoi{10.1111/j.1365-2966.2012.21181.x}

\bibitem[{{Zilh{\~a}o} {et~al.}(2015){Zilh{\~a}o}, {Noble}, {Campanelli}, \&
  {Zlochower}}]{Zilhao2015}
{Zilh{\~a}o}, M., {Noble}, S.~C., {Campanelli}, M., \& {Zlochower}, Y. 2015,
  \prd, 91, 024034, \dodoi{10.1103/PhysRevD.91.024034}

\bibitem[{{Zrake} {et~al.}(2021){Zrake}, {Tiede}, {MacFadyen}, \&
  {Haiman}}]{Zrake2021}
{Zrake}, J., {Tiede}, C., {MacFadyen}, A., \& {Haiman}, Z. 2021, \apjl, 909,
  L13, \dodoi{10.3847/2041-8213/abdd1c}

\end{thebibliography}
\bibliographystyle{aasjournal}



\end{document}